\documentclass{easychair}
\usepackage{amssymb}
\usepackage{tabularx}
\usepackage{multirow}
\usepackage{tikz}
\usepackage{cleveref}
\usepackage{float}
\usepackage{amsmath}
\usepackage{algpseudocode}
\usepackage{algorithm2e}

\usetikzlibrary{positioning}

\title{On the strong metric dimension of composed graphs}
\author{Marcel Wagner \and Yannick Schmitz \and Egon Wanke}
\institute{Heinrich-Heine-Universit\"at D\"usseldorf, Germany\\ \email{ marcel.wagner@hhu.de, yannick.schmitz@hhu.de, egon.wanke@hhu.de}}
\newtheorem{theorem}{Theorem}
\newtheorem{definition}{Definition}
\newtheorem{notation}{Notation}

\newtheorem{lemma}{Lemma}

\newtheorem{observation}{Observation}
\newtheorem{prerequisite}{Prerequisite}

\authorrunning{Wagner, Schmitz, Wanke}
\titlerunning{Strong metric dimension of composed graphs}

\newcommand{\capfont}[1]{\textsf{\textit{#1}}}

\newcommand{\MERGE}[2]{{\, \circ_{(#1 \to #2)} \,}}
\newcommand{\MD}[2]{{\text{MD}(\,#1,\,#2\,)}}
\newcommand{\MVC}[1]{{\text{VC}(\,#1\,)}}
\newcommand{\SR}[1]{{#1}_{\rm SR}}

\newcommand{\XVC}[2]{{\overline{\text{VC}}(\,#1,\,#2\,)}}
\newcommand{\union}{\hbox{\small$\cup$}\,}
\newcommand{\join}{\hbox{$\times$}\,}
\newcommand{\sv}[1]{{\rm s}(#1)}
\newcommand{\vertex}[1]{{\rm vertex}(#1)}

\newcommand{\node}[1]{{\rm node}(#1)}

\begin{document}

\maketitle

\begin{abstract}
Two vertices $u$ and $v$ of an undirected graph $G$ are strongly resolved by a vertex $w$ if there is a shortest path between $w$ and $u$ containing $v$ or a shortest path between $w$ and $v$ containing $u$.
A vertex set $R$ is a strong resolving set for $G$ if for each pair of vertices there is a vertex in $R$ that strongly resolves them.
The strong metric dimension of $G$ is the size of a minimum strong resolving set for $G$.
We show that a minimum strong resolving set for an undirected graph $G$ can be computed efficiently if and only if a minimum strong resolving set for each biconnected component of $G$ can be computed efficiently.
\end{abstract}

%%%%%%%%%%%%%%%%%%%%%%%%%%%%%%%%%%%%%%%%%%%%%%%%%%%%%%%%%%%%
% Section 1
%%%%%%%%%%%%%%%%%%%%%%%%%%%%%%%%%%%%%%%%%%%%%%%%%%%%%%%%%%%%

\section{Introduction}
\label{section1}

In this paper we consider the {\em strong metric dimension} introduced by Seb{\"{o}} and Tannier in \cite{ST04}. The strong metric dimension is a variant of the original {\em metric dimension} (which we simply call metric dimension) which is the smallest number $k$ of vertices from which the vector of distances to every vertex in the graph is unique. Here the distance between two vertices is the number of edges on a shortest path. The $k$-dimensional distance vectors of the vertices can be viewed as their positions in a $k$-dimensional space whose structure is defined by the graph.

The metric dimension has been introduced by Slater in \cite{Sla75} and \cite{Sla88} and independently by Harary and Melter in \cite{HM76}. There are numerous research reports on the analysis of the metric dimension of graphs. Determining whether the metric dimension of a given graph is less than a given integer has been shown to be NP-complete by a reduction from {\sc3-SAT} \cite{KRR96} and {\sc 3-Dimensional Matching} \cite{GJ79}. It is NP-complete for general graphs, planar graphs \cite{DPSL17}, even for those with maximum degree 6, and Gabriel unit disk graphs \cite{HW12}. There are several algorithms for computing the metric dimension in polynomial time for special classes of graphs, as for example for trees \cite{CEJO00,KRR96}, wheels \cite{HMPSCP05}, grid graphs \cite{MT84}, $k$-regular bipartite graphs \cite{SBSSB11}, amalgamation of cycles \cite{IBSS10}, outerplanar graphs \cite{DPSL17}, cactus block graphs \cite{hoffmann2016linear}, chain graphs \cite{fernau2015computing}, and graphs with bounded extended biconnected components \cite{VHW19}.

The strong metric dimension of a graph $G$, in contrast to the metric dimension, is the size of a smallest set $R$ of vertices with the following property. For each pair of two distinct vertices $u$ and $v$ in $G$, there is a vertex $w\in R$ such that there is a shortest path between $w$ and $u$ that contains $v$ or a shortest path between $w$ and $v$ that contains $u$. Such a set $R$ is called a {\em strong resolving set} for $G$. Since in both cases the distance between $w$ and $u$ is different from the distance between $w$ and $v$, a strong resolving set for $G$ is always a resolving set for $G$ and thus the strong metric dimension is always greater than or equal to the metric dimension. However, if we again calculate the distance vectors
$\vec{u} = (d_G(u,w_1),\ldots,d_G(u,w_k))$ for vertices $u$ to the $k$ vertices $w_1,\ldots, w_k$ of a strong resolving set $R$, then there are significant advantages in contrast to the metric dimension when navigating through the graph. The distance between two vertices $u$ and $v$ is the maximum difference between $d_G(u,w_i)$ and $d_G(v,w_i)$ for $i=1,\ldots,k$, that is, $$d_G(u,v) \, = \, \max_{i=1}^k \vert d_G(u,w_i)-d_G(v,w_i) \vert.$$ To navigate from a vertex $u$ to a vertex $v$ in graph $G$, we can now simply determine a neighbour $u'$ of $u$ on a shortest path between $u$ and $v$. This is a neighbour $u'$ of $u$ with $d_G(u',v) = d_G(u,v) -1$.

Determining whether the strong metric dimension of a given graph is less than a given integer $k$ is NP-complete \cite{OP07}, like it is the case for the metric dimension. Computing the strong metric dimension also has been extensively studied for different graph classes, see for example \cite{LZZ20}, \cite{KYR16}, \cite{Man15}, \cite{Kuz20}, \cite{FM19} and \cite{WK18}.

In this paper we show that an efficient computation of the strong metric dimension for a graph $G$ can be reduced to an efficient computation of the biconnected components of $G$. That is, we consider a composition mechanism that connects two graphs $G_1$ and $G_2$ by identifying a vertex from $G_1$ with a vertex from $G_2$ in the disjoint union of $G_1$ and $G_2$.
With this composition mechanism, a graph can be assembled from its biconnected components.
Computing the strong metric dimension for graphs obtained by join operations, like the Cartesian product, the strong product and the corona product, has also been studied by other authors, see for example \cite{KYR13} and \cite{RYK14}. We demonstrate the power of our approach by three examples. We show that the strong metric dimension for graphs in which the biconnected components are circles or co-graphs can be computed in linear time.

%%%%%%%%%%%%%%%%%%%%%%%%%%%%%%%%%%%%%%%%%%%%%%%%%%%%%%%%%%%%
% Section 2
%%%%%%%%%%%%%%%%%%%%%%%%%%%%%%%%%%%%%%%%%%%%%%%%%%%%%%%%%%%%

\section{Strong metric dimension}
\label{section2}

We consider undirected, connected, and finite graphs $G=(V,E)$, where $V$ is the set of vertices and $E \subseteq \{\{u,v\}\ |\ u,v \in V, u\neq v\}$ is the set of edges. Two distinct vertices $u,v \in V$ of $G$ are {\em strongly resolved} by a vertex $w \in V$ if there is a shortest path between $w$ and $u$ that contains $v$ or a shortest path between $w$ and $v$ that contains $u$. The length of a path is the number of edges. A vertex set $R \subseteq V$ is a {\em strong resolving set} for $G$ if for each pair of vertices $u,v \in V \setminus R, u \neq v$ there is a vertex $w \in R$ such that $u$ and $v$ are strongly resolved by $w$. The {\em strong metric dimension} of graph $G$ is the size of a smallest strong resolving set for $G$.

Ollermann and Peters-Fransen showed in \cite{OP07} that finding a strong resolving set of $G$ is equivalent to finding a vertex cover of the so-called {\em strong resolving graph} $\SR{G}$ of $G$, defined as follows. For a vertex $u \in V$ let $N_G(u) = \{ v \,\vert\, \{u,v\} \in E\}$ and $N_G[u] = N_G(u) \cup \{u\}$ be the open and closed neighbourhoods of $u$, respectively. For two vertices $u,v \in V$ let $d_G(u,v)$ be the distance between $u$ and $v$ in $G$, that is, the number of edges of a shortest path between $u$ and $v$. We say a vertex $u \in V$ is {\em maximally distant} from a vertex $w$ if there is no vertex $v \in N_G(u)$ in the neighbourhood of $u$ with $d_G(v,w) > d_G(u,w)$.

\begin{figure}[H]
\center
\includegraphics[width=240pt]{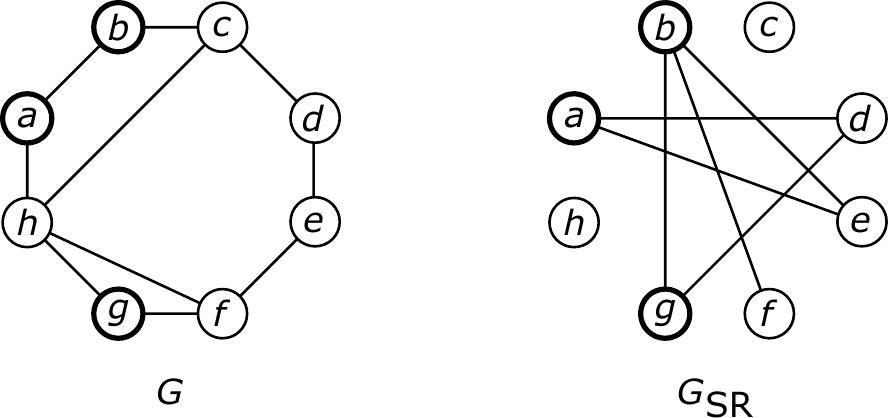}
\caption{A Graph $G$ and its strong resolving graph $\SR{G}$. The vertex set $\{\capfont{a,b,g}\}$ is a minimum vertex cover for $\SR{G}$ and a minimum strong resolving set for $G$.}
\label{figure1}
\end{figure}

The vertices of the strong resolving graph $\SR{G}$ are the vertices of $G$. There is an edge between two vertices $u$ and $v$ in $\SR{G}$ if and only if $u$ is maximally distant from $v$ and $v$ is maximally distant from $u$. In this case we also say that $u$ and $v$ are {\em mutually maximally distant}. It is easy to see that each strong resolving set for $G$ must contain at least one of two vertices that are mutually maximally distant. Also each set of vertices that contains at least one of two vertices that are mutually maximally distant is a strong resolving set for $G$. It follows that a strong resolving set for $G$ is a vertex cover for $\SR{G}$ and vice versa. See \Cref{figure1} for an example.

%%%%%%%%%%%%%%%%%%%%%%%%%%%%%%%%%%%%%%%%%%%%%%%%%%%%%%%%%%%%
% Figure 1
%%%%%%%%%%%%%%%%%%%%%%%%%%%%%%%%%%%%%%%%%%%%%%%%%%%%%%%%%%%%

%%%%%%%%%%%%%%%%%%%%%%%%%%%%%%%%%%%%%%%%%%%%%%%%%%%%%%%%%%%%
% Section 3
%%%%%%%%%%%%%%%%%%%%%%%%%%%%%%%%%%%%%%%%%%%%%%%%%%%%%%%%%%%%

\section{Composing graphs}
\label{section3}

The graphs we consider arise from attaching child graphs $G_1,\ldots,G_k$ to a parent graph $H$. This attachment is performed by merging vertices $u_1,\ldots,u_k$ of $G_1,\ldots,G_k$ with vertices $v_1,\ldots, v_k$ from $H$ in the disjoint union of the graphs $G_1,\ldots,G_k$, and $H$.

%%%%%%%%%%%%%%%%%%%%%%%%%%%%%%%%%%%%%%%%%%%%%%%%%%%%%%%%%%%%
% Definition 1
%%%%%%%%%%%%%%%%%%%%%%%%%%%%%%%%%%%%%%%%%%%%%%%%%%%%%%%%%%%%

\begin{definition}
\label{definition1}
Let $G_i=(V_i,E_i)$, $1 \leq i \leq k$, and $H=(V_H,E_H)$ be $k+1$ graphs, $u_i \in V_i$ for $i=1,\ldots,k$, and $v_1,\ldots,v_k \in V_H$. Let $G_{1,\ldots,k}$ be the disjoint union of $G_1,\ldots,G_k$. That is, $G_{1,\ldots,k}$ has vertex set $\cup_{i=1}^k V_i$ and edge set $\cup_{i=1}^k E_i$.

Then graph
$$G_{1,\ldots,k} \MERGE{u_1,\ldots,u_k}{v_1,\ldots,v_k} H$$
is defined by vertex set
$$
(V_1 \cup \cdots \cup V_k \cup V_H) \setminus \{u_1,\ldots,u_k\}
$$
and edge set
$$
\begin{array}{lll}
(         & E_1 \cup \cdots \cup E_k \cup E_H & \\
          & \cup \, \{ \{w,v_i\} \,\vert\, w \in N_{G_i}(u_i), 1 \leq i \leq k \} & ) \\
\setminus & \{ \{w,u_i\} \,\vert\, w \in N_{G_i}(u_i), 1 \leq i \leq k \}. & 
\end{array}
$$
\end{definition}

%%%%%%%%%%%%%%%%%%%%%%%%%%%%%%%%%%%%%%%%%%%%%%%%%%%%%%%%%%%%

\begin{figure}[H]
\center
\includegraphics[width=300pt]{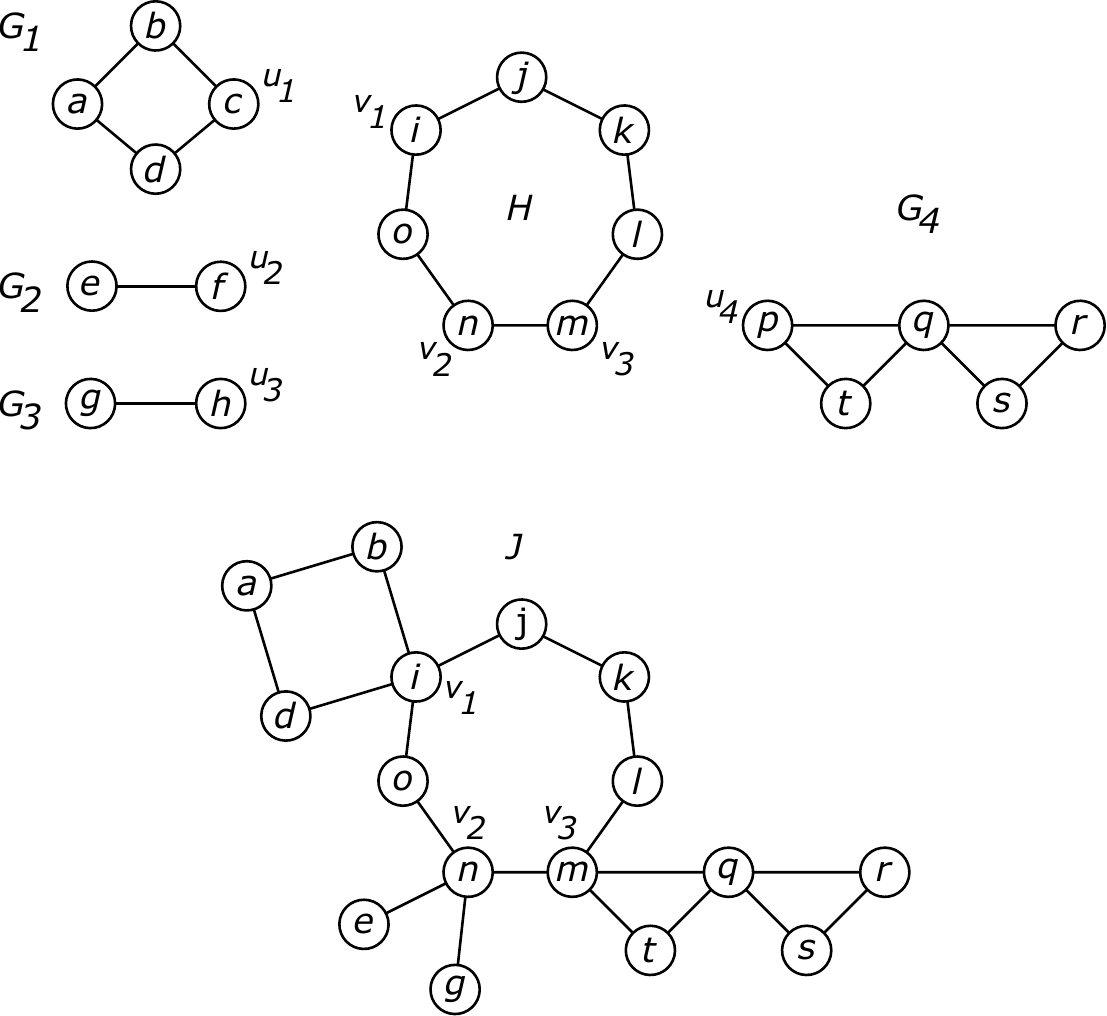}
\caption{Five graphs $G_1$, $G_2$, $G_3$, $G_4$, $H$, and the graph $J = G_{1,2,3,4} \MERGE{u_1,u_2,u_3,u_4}{v_1,v_2,v_2,v_3} H$ created by the composition as defined in \Cref{definition1}.}
\label{figure2}
\end{figure}

The graph $G_{1,\ldots,k} \MERGE{u_1,\ldots,u_k}{v_1,\ldots,v_k} H$ is formed by the disjoint union of the $k$ graphs $G_1,\ldots,G_k$ and graph $H$ without the vertices $u_1,\ldots,u_k$ and their incident edges, in which for $i=1,\ldots,k$ the neighbours of vertex $u_i$ in $G_i$ are connected to vertex $v_i$. \Cref{figure2} shows an example of the $\MERGE{u_1,\ldots,u_k}{v_1,\ldots,v_k}$ operation. For all further discussions, we only consider the case in that all graphs $G_1,\ldots,G_k,H$ are vertex disjoint, connected, and have at least two vertices. The vertices $v_1,\ldots,v_k$ of $H$ do not need to be distinct.

%%%%%%%%%%%%%%%%%%%%%%%%%%%%%%%%%%%%%%%%%%%%%%%%%%%%%%%%%%%%
% Figure 2
%%%%%%%%%%%%%%%%%%%%%%%%%%%%%%%%%%%%%%%%%%%%%%%%%%%%%%%%%%%%

%%%%%%%%%%%%%%%%%%%%%%%%%%%%%%%%%%%%%%%%%%%%%%%%%%%%%%%%%%%%
% Section 4
%%%%%%%%%%%%%%%%%%%%%%%%%%%%%%%%%%%%%%%%%%%%%%%%%%%%%%%%%%%%

\section{The strong resolving graph}
\label{section4}

To compute a strong resolving set for a composed graph $J = G_{1,\ldots,k} \MERGE{u_1,\ldots,u_k}{v_1,\ldots,v_k} H$, we first determine the edge set of the strong resolving graph $\SR{J}$.

%%%%%%%%%%%%%%%%%%%%%%%%%%%%%%%%%%%%%%%%%%%%%%%%%%%%%%%%%%%%
% Definition 2
%%%%%%%%%%%%%%%%%%%%%%%%%%%%%%%%%%%%%%%%%%%%%%%%%%%%%%%%%%%%

\begin{definition}
\label{definition2}
For a connected graph $G=(V,E)$ and a vertex $u \in V$, let $\MD{G}{u}$ be the set of vertices that are maximally distant from vertex $u$ in $G$. 
\end{definition}

The following prerequisite is used in each of the following lemmas and theorems.

%%%%%%%%%%%%%%%%%%%%%%%%%%%%%%%%%%%%%%%%%%%%%%%%%%%%%%%%%%%%
% Prerequisite 1
%%%%%%%%%%%%%%%%%%%%%%%%%%%%%%%%%%%%%%%%%%%%%%%%%%%%%%%%%%%%

\begin{prerequisite}
\label{prerequisite1}
Let $G_i=(V_i,E_i)$, $1 \leq i \leq k$, and $H=(V_H,E_H)$ be $k+1$ vertex disjoint, connected graphs, $|V_i| \geq 2$, $u_i \in V_i$ for $i=1,\ldots,k$, $|V_H| \geq 2$, $v_1,\ldots,v_k \in V_H$, let $G_{1,\ldots,k}$ be the disjoint union of $G_1,\ldots,G_k$, and let $$J = (V_J,E_J) = G_{1,\ldots,k} \MERGE{u_1,\ldots,u_k}{v_1,\ldots,v_k} H.$$ The vertices $v_1,\ldots,v_k$ of $H$ do not need to be distinct.
\end{prerequisite}

%%%%%%%%%%%%%%%%%%%%%%%%%%%%%%%%%%%%%%%%%%%%%%%%%%%%%%%%%%%%
% Lemma 1
%%%%%%%%%%%%%%%%%%%%%%%%%%%%%%%%%%%%%%%%%%%%%%%%%%%%%%%%%%%%

\begin{lemma}
\label{lemma1}
Let \Cref{prerequisite1} be given. Then the vertices $v_1,\ldots,v_k$ have no incident edges in $\SR{J}$.

\begin{proof}
The vertices $v_1,\ldots,v_k$ are separation vertices in $J$, because all graphs $G_i$, $1 \leq i \leq k$, and $H$ are connected and have at least two vertices. Separation vertices are not maximally distant from any vertex.
\end{proof}
\end{lemma}

%%%%%%%%%%%%%%%%%%%%%%%%%%%%%%%%%%%%%%%%%%%%%%%%%%%%%%%%%%%%
% Lemma 2
%%%%%%%%%%%%%%%%%%%%%%%%%%%%%%%%%%%%%%%%%%%%%%%%%%%%%%%%%%%%

\begin{lemma}
\label{lemma2}
Let \Cref{prerequisite1} be given. Let $w \in V_H$, then
$$
\begin{array}{ll}
\MD{J}{w} = & 
\left(\begin{array}{ll} & \MD{H}{w} \setminus \{v_1,\ldots,v_k\} \\ \cup & \bigcup_{1 \leq i \leq k} \MD{G_i}{u_i}
\end{array}\right).
\end{array}
$$

\begin{proof}
This follows from the fact that each shortest path in $J$ between $w$ and a vertex of $G_i$ for $1 \leq i \leq k$ passes vertex $v_i$ and each shortest path in $J$ between $w$ and a vertex of $H$ does not pass a vertex outside of $H$. Separation vertices are not maximally distant from any vertex, see also \Cref{lemma1}.
\end{proof}
\end{lemma}

In \Cref{figure2} the vertices $\{\capfont{a,o,e,g,t,s,r}\}$ are maximally distant from vertex $\capfont{l}$ in $J$. 

%%%%%%%%%%%%%%%%%%%%%%%%%%%%%%%%%%%%%%%%%%%%%%%%%%%%%%%%%%%%
% Lemma 3
%%%%%%%%%%%%%%%%%%%%%%%%%%%%%%%%%%%%%%%%%%%%%%%%%%%%%%%%%%%%

\begin{lemma}
\label{lemma3}
Let \Cref{prerequisite1} be given. Let $u'_j \in V_j \setminus \{u_j\}$ for some $j$, $1 \leq j \leq k$. Then
$$
\begin{array}{ll}
\MD{J}{u'_j} = &
\left(\begin{array}{ll} & \MD{G_j}{u'_j} \setminus \{u_j\} \\ \cup & \MD{H}{v_j} \setminus \{v_1,\ldots,v_k\} \\ \cup & \bigcup_{1 \leq i \leq k, i \not=j} \MD{G_i}{u_i} \end{array}\right).
\end{array}$$

\begin{proof}
This follows from the fact that each shortest path in $J$ between $u'_j$ and a vertex of $H$ or $G_i$ for $i \not= j$ passes vertex $v_j$ and each shortest path in $J$ between $u'_j$ and a vertex of $(V_j \setminus \{u_j\}) \cup \{v_j\}$ do not pass a vertex outside of $(V_j \setminus \{u_j\}) \cup \{v_j\}$. Separation vertices are not maximally distant from any vertex, see also \Cref{lemma1}. Note that vertex $u_j$ can be maximally distant from $u'_j$ in $G_j$, but is not a vertex of $J$. This is why we need to remove $u_j$ as well from $\MD{G_j}{u'_j}$.
\end{proof}
\end{lemma}

%%%%%%%%%%%%%%%%%%%%%%%%%%%%%%%%%%%%%%%%%%%%%%%%%%%%%%%%%%%%

In \Cref{figure2} the vertices $\{\capfont{l,e,g,t,s,r}\}$ are maximally distant from vertex $\capfont{a}$ in $J$. Note that vertex $\capfont{c}$ is also maximally distant from vertex $\capfont{a}$ in $G_1$, but is excluded from $\MD{G_1}{\capfont{a}}$.

The next lemmas characterise the edges of $\SR{J}$ between vertices of $V_i \setminus \{u_i\}$ and vertices of $V_j \setminus \{u_j\}$ for $i\neq j$, and the edges between vertices of $V_i \setminus \{u_i\}$ and vertices of $V_H$.

%%%%%%%%%%%%%%%%%%%%%%%%%%%%%%%%%%%%%%%%%%%%%%%%%%%%%%%%%%%%
% Lemma 4
%%%%%%%%%%%%%%%%%%%%%%%%%%%%%%%%%%%%%%%%%%%%%%%%%%%%%%%%%%%%

\begin{lemma}
\label{lemma4}
Let \Cref{prerequisite1} be given. Then for each vertex $u'_i \in V_i \setminus \{u_i\}$ for some $i$, $1 \leq i \leq k$, and each vertex $v' \in V_J \setminus V_i$, the following statements hold true.
\begin{enumerate}
\item If $u'_i$ is maximally distant from $u_i$ in $G_i$, or equivalently maximally distant from $v_i$ in $J$, then $u'_i$ is maximally distant from $v'$ in $J$.
\item If $v'$ is maximally distant from $v_i$ in $J$, then $v'$ is maximally distant from $u'_i$ in $J$.
\end{enumerate}

\begin{proof}
This follows again from the fact that each shortest path in $J$ between $u'_i$ and $v'$ passes vertex $v_i$, and thus  $d_J(u'_i,v') = d_J(u'_i,v_i) + d_J(v_i,v')$.
\end{proof}
\end{lemma}

In \Cref{figure2} vertex $\capfont{a}$ is maximally distant from vertex $\capfont{c}$ in $G_1$, thus vertex $\capfont{a}$ is maximally distant from all vertices except for $\capfont{b}$ and $\capfont{d}$ in $J$. Also, vertex $\capfont{e}$ is maximally distant from vertex $\capfont{f}$ in $G_2$, thus the vertices $\capfont{a}$ and $\capfont{e}$ are mutually maximally distant in $J$.

%%%%%%%%%%%%%%%%%%%%%%%%%%%%%%%%%%%%%%%%%%%%%%%%%%%%%%%%%%%%
% Lemma 5
%%%%%%%%%%%%%%%%%%%%%%%%%%%%%%%%%%%%%%%%%%%%%%%%%%%%%%%%%%%%

\begin{lemma}
\label{lemma5}
Let \Cref{prerequisite1} be given. Two vertices $u'_i \in V_i \setminus \{u_i\}$ and $u'_j \in V_j \setminus \{u_j\}$ for $i \not= j$ are mutually maximally distant in $J$ if and only if $u'_i$ is maximally distant from $u_i$ in $G_i$ and $u'_j$ is maximally distant from $u_j$ in $G_j$.

\begin{proof}
~

"$\Rightarrow$"
Let $u'_i$ be maximally distant from $u_i$ in $G_i$ and $u'_j$ be maximally distant from $u_j$ in $G_j$. By \Cref{lemma4}, $u'_i$ is maximally distant from each vertex of $V_J\setminus V_i$ and $u'_j$ is maximally distant to each vertex of $V_J \setminus V_j$. Thus $u'_i$ and $u'_j$ are mutually maximally distant in $J$ and $\{u'_i,u'_j\}$ is an edge in $\SR{J}$.

"$\Leftarrow$" Since each path between $u'_i$ and $u'_j$ in $J$ passes vertex $u_i$ (and vertex $u_j$), the following statement holds true. If $u'_i$ and $u'_j$ are mutually maximally distant in $J$, then $u'_i$ is maximally distant from $u_i$ in $G_i$ and $u'_j$ is maximally distant from $u_j$ in $G_j$.
\end{proof}
\end{lemma}

%%%%%%%%%%%%%%%%%%%%%%%%%%%%%%%%%%%%%%%%%%%%%%%%%%%%%%%%%%%%

\Cref{lemma5} identifies the edges of $\SR{J}$ between $u'_i \in V_i$ and $u'_j \in V_j$ for $i\neq j$.

Next we identify the edges of $\SR{J}$ between two vertices of $V_i \setminus \{u_i\}$ and between two vertices of $V_H \setminus \{v_1,\ldots,v_k\}$.

%%%%%%%%%%%%%%%%%%%%%%%%%%%%%%%%%%%%%%%%%%%%%%%%%%%%%%%%%%%%
% Lemma 6
%%%%%%%%%%%%%%%%%%%%%%%%%%%%%%%%%%%%%%%%%%%%%%%%%%%%%%%%%%%%

\begin{lemma}
\label{lemma6}
Let \Cref{prerequisite1} be given.
\begin{enumerate}
\item
Two vertices $u'_i,u''_i \in V_i\setminus\{u_i\}$ are mutually maximally distant in $J$ if and only if they are mutually maximally distant in $G_i$.
\item
Two vertices $v',v'' \in V_H\setminus\{v_1,\ldots,v_k\}$ are mutually maximally distant in $J$ if and only if they are mutually maximally distant in $H$.
\end{enumerate}

\begin{proof}
The statements follow from the facts that each shortest path between $u'_i$ and $u''_i$ in $J$ does not pass a vertex of $V_J \setminus (V_i \cup \{v_i\})$ and each shortest path between $v'$ and $v''$ in $J$ does not pass a vertex of $V_1 \cup \ldots \cup V_k$.
\end{proof}
\end{lemma}

%%%%%%%%%%%%%%%%%%%%%%%%%%%%%%%%%%%%%%%%%%%%%%%%%%%%%%%%%%%%

The following theorem follows from \Cref{lemma5} and \Cref{lemma6} and characterises all edges of $\SR{J}$. \Cref{figure3} shows the strong resolving graph $\SR{J}$ of $J$ from \Cref{figure2}.

%%%%%%%%%%%%%%%%%%%%%%%%%%%%%%%%%%%%%%%%%%%%%%%%%%%%%%%%%%%%
% Theorem 1
%%%%%%%%%%%%%%%%%%%%%%%%%%%%%%%%%%%%%%%%%%%%%%%%%%%%%%%%%%%%

\begin{theorem}
\label{theorem1}
Let \Cref{prerequisite1} be given. The strong resolving graph $\SR{J}$ has an edge $\{w_1,w_2\}$ if and only if $w_1, w_2 \notin \{v_1,\ldots,v_k\}$ and
\begin{enumerate}
\item $w_1, w_2 \in V_i$ for some $i$, $1 \leq i \leq k$, and $w_1$ and $w_2$ are  mutually maximally distant in $G_i$,
\item $w_1, w_2 \in V_H$ and $w_1$ and $w_2$ are mutually maximally distant in $H$,
\item $w_1 \in V_i$ and $w_2 \in V_j$ for some $i,j$, $1 \leq i < j \leq k$, and $w_1$ is maximally distant from $u_i$ in $G_i$ and $w_2$ is maximally distant from $u_j$ in $G_j$, or
\item if $w_1 \in V_i$ for some $i$, $1 \leq i \leq k$, $w_2 \in V_H$, $w_1$ is maximally distant from $u_i$ in $G_i$ and $w_2$ is maximally distant from $v_i$ in $H$.
\end{enumerate}
\end{theorem}

%%%%%%%%%%%%%%%%%%%%%%%%%%%%%%%%%%%%%%%%%%%%%%%%%%%%%%%%%%%%
% Section 5
%%%%%%%%%%%%%%%%%%%%%%%%%%%%%%%%%%%%%%%%%%%%%%%%%%%%%%%%%%%%

\section{A minimum vertex cover}
\label{section5}

%%%%%%%%%%%%%%%%%%%%%%%%%%%%%%%%%%%%%%%%%%%%%%%%%%%%%%%%%%%%

\begin{figure}[H]
\center
\includegraphics[width=344pt]{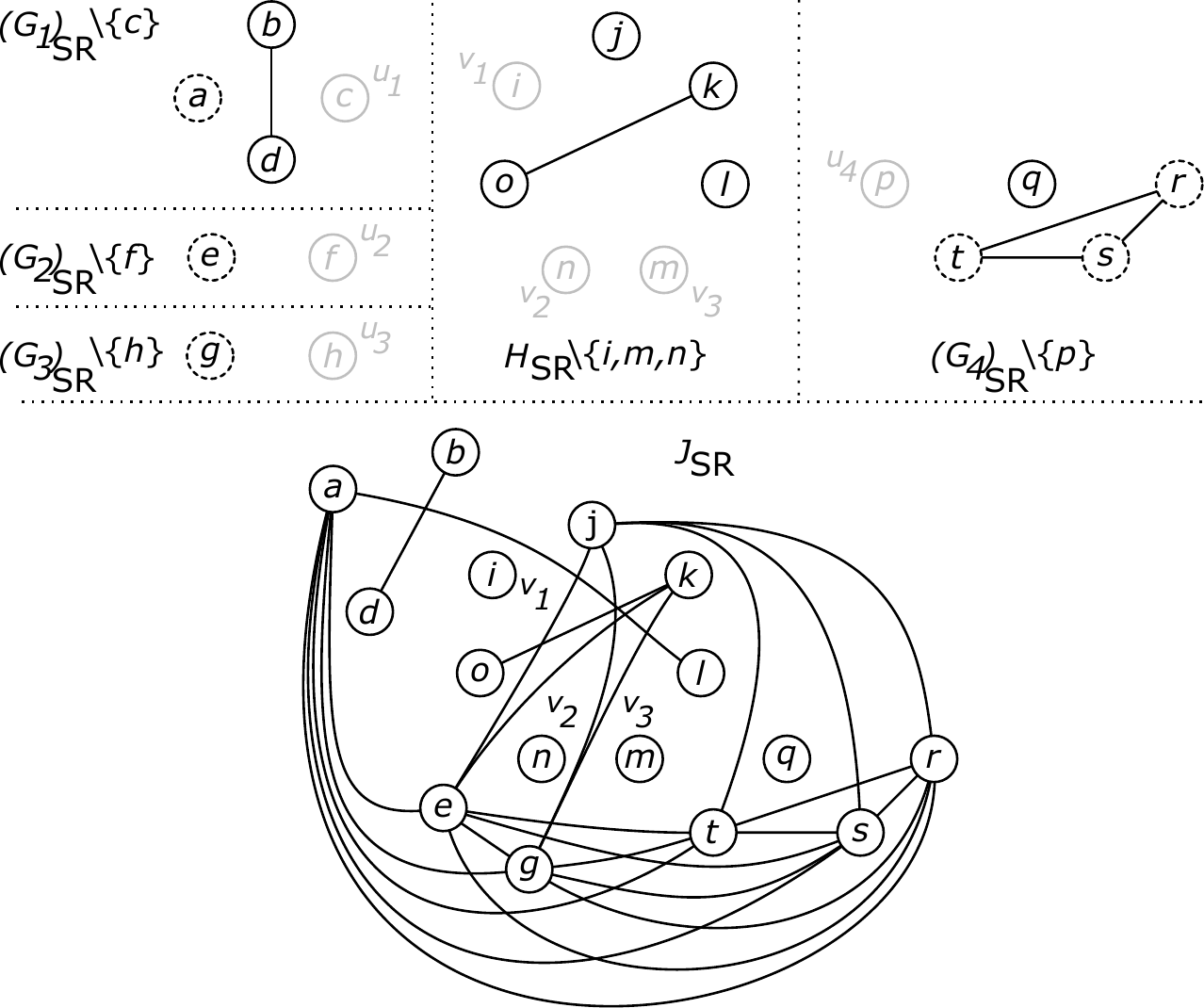}
\caption{The graphs $\SR{(G_1)} \setminus \{\capfont{c}\}$, $\SR{(G_2)} \setminus \{\capfont{f}\}$, $\SR{(G_3)} \setminus \{\capfont{h}\}$, $\SR{(G_4)} \setminus \{\capfont{p}\}$, $\SR{H} \setminus \{\capfont{i,m,n}\}$, and $\SR{J}$, where $G_1$, $G_2$, $G_3$, $G_4$, $H$, and $J$ are from \Cref{figure2}. The dashed vertices in $\SR{(G_1)}$, $\SR{(G_2)}$, $\SR{(G_3)}$, and $\SR{(G_4)}$ are the vertices which are maximally distant from $u_1$, $u_2$, $u_3$, and $u_4$ in $G_1$, $G_2$, $G_3$, and $G_4$, respectively. The grey vertices are the vertices that are merged to separation vertices of $J$. These vertices are not taken into account when computing minimum strongly resolving sets for $J$.}
\label{figure3}
\end{figure}

\Cref{theorem1} characterises the edges in the strong resolving graph $\SR{J}$ as follows, see \Cref{figure3}.

%%%%%%%%%%%%%%%%%%%%%%%%%%%%%%%%%%%%%%%%%%%%%%%%%%%%%%%%%%%%
% Figure 3
%%%%%%%%%%%%%%%%%%%%%%%%%%%%%%%%%%%%%%%%%%%%%%%%%%%%%%%%%%%%

%%%%%%%%%%%%%%%%%%%%%%%%%%%%%%%%%%%%%%%%%%%%%%%%%%%%%%%%%%%%

\begin{itemize}
\item The edges considered in Case 1 are the edges of $\SR{(G_i)} \setminus \{u_i\}$.
\item The edges considered in Case 2 are the edges of $\SR{H} \setminus \{v_1,\ldots,v_k\}$.
\item The edges considered in Case 3 are the edges of a complete $k$-partite graph $K_{n_1,\ldots,n_k}$, $n_i =  \vert\MD{G_i}{u_i}\vert$, with vertex set
$$\bigcup_{1 \leq i \leq k} \MD{G_i}{u_i}$$ and edge set
$$\{\{w_1,w_2\} \,\vert\, w_1, \in \MD{G_i}{u_i}, w_2, \in \MD{G_j}{u_j}, 1 \leq i < j \leq k\},$$.
\item The edges considered in Case 4 are the edges of $k$ complete bipartite graphs $K_{n_i,m_i}$, $1 \leq i \leq k$, $n_i =  \vert\MD{G_i}{u_i}\vert$, $m_i=\vert\MD{H}{v_i} \setminus \{v_1,\ldots,v_k\}\vert$ with vertex set
$$\MD{G_i}{u_i} \, \cup \, (\,\MD{H}{v_i} \setminus \{v_1,\ldots,v_k\}\,)$$ and edge set
$$\{\{w_1,w_2\} \,\vert\, w_1, \in \MD{G_i}{u_i}, w_2, \in \MD{H}{v_i} \setminus \{v_1,\ldots,v_k\}\}.$$
\end{itemize}

As mentioned in \Cref{section2}, a vertex set is a strong resolving set for $J$ if and only if it is a vertex cover for $\SR{J}$. The strong resolving graph $\SR{J}$ contains the $k$-partite graph $K_{n_1,\ldots,n_k}$ and the $k$ bipartite graphs $K_{n_i,m_i}$, $1 \leq i \leq k$, as subgraphs. That is, each vertex cover $U$ for $\SR{J}$ contains for some $j$, $1 \leq j \leq k$, all vertices of the vertex sets $\MD{G_i}{u_i}$, $1\leq i\leq k$, $i\neq j$, and additionally either all vertices of $\MD{G_j}{u_j}$ or all vertices of $\MD{H}{v_j} \setminus \{v_1,\ldots,v_k\}$.

To compute the size of a minimum vertex cover of $\SR{J}$, we use the following notation of a restricted vertex cover.

%%%%%%%%%%%%%%%%%%%%%%%%%%%%%%%%%%%%%%%%%%%%%%%%%%%%%%%%%%%%
% Notation 1
%%%%%%%%%%%%%%%%%%%%%%%%%%%%%%%%%%%%%%%%%%%%%%%%%%%%%%%%%%%%

\begin{notation}
\label{notation1}
For a graph $G=(V,E)$ let $$\MVC{G}$$ be a minimum vertex cover for $G$.
For a graph $G=(V,E)$ and a vertex set $M \subseteq V$ let $$\XVC{G}{M}$$ be a vertex set of minimum size that contains all vertices of $M$ and which is a vertex cover for $G$.
\end{notation}

%%%%%%%%%%%%%%%%%%%%%%%%%%%%%%%%%%%%%%%%%%%%%%%%%%%%%%%%%%%%
% Lemma 7
%%%%%%%%%%%%%%%%%%%%%%%%%%%%%%%%%%%%%%%%%%%%%%%%%%%%%%%%%%%%

With the help of \Cref{notation1}, a minimum vertex cover for $\SR{J}$ can now easily be specified.

\begin{lemma}
\label{lemma7}
Let \Cref{prerequisite1} be given. Let
$$U_0 = \left(\begin{array}{ll}
     & \MVC{\SR{H} \setminus \{v_1,\ldots,v_k\}} \\
\cup & \bigcup_{1 \leq i \leq k} \XVC{\SR{(G_i)} \setminus \{u_i\}}{\MD{G_i}{u_i}}
\end{array} \right)
$$
and
$$U_j = \left(\begin{array}{ll}
     & \XVC{\SR{H} \setminus \{v_1,\ldots,v_k\}}{\MD{H}{v_j} \setminus \{v_1,\ldots,v_k\}} \\
\cup & \MVC{\SR{(G_j)} \setminus \{u_j\}} \\
\cup & \bigcup_{1 \leq i \leq k, i\not=j} \XVC{\SR{(G_i)} \setminus \{u_i\}}{\MD{G_i}{u_i}}
\end{array} \right)
$$
for $j=1,\ldots,k$.

Each vertex sets $U_0,U_1,\ldots,U_k$ is a vertex cover for $\SR{J}$, where at least one of them is a minimum vertex cover $\SR{J}$.

\begin{proof}
As mentioned above, $\SR{J}$ contains a complete $k$-partite subgraph with vertex set $$\MD{G_1}{u_1} \, \cup \, \cdots \, \cup \, \MD{G_k}{u_k}$$ and $k$ bipartite subgraphs with vertex sets $$\MD{G_i}{u_i} \, \cup \, (\MD{H}{u_i} \setminus \{u_1,\ldots ,u_k\})$$ for $i=1,\ldots,k$. Each vertex cover of $\SR{J}$ must therefore contain all vertices from all sets $\MD{G_1}{u_1}, \ldots, \MD{G_k}{u_k}$ except for one of these sets $\MD{G_j}{u_j}$, $1\leq j\leq k$, and must additionally contain either all vertices of $\MD{G_j}{u_j}$ or all vertices of $\MD{H}{v_j} \setminus \{v_1,\ldots,v_k\}$. The edges in $\SR{J}$ which are not incident with the vertices of the selected sets must of course also be covered by a vertex cover. These $k+1$ cases are treated by considering the sets $U_0, U_1,\ldots,U_k$.
\end{proof}
\end{lemma}

%%%%%%%%%%%%%%%%%%%%%%%%%%%%%%%%%%%%%%%%%%%%%%%%%%%%%%%%%%%%

\begin{figure}[htb]
\center
\includegraphics[width=195pt]{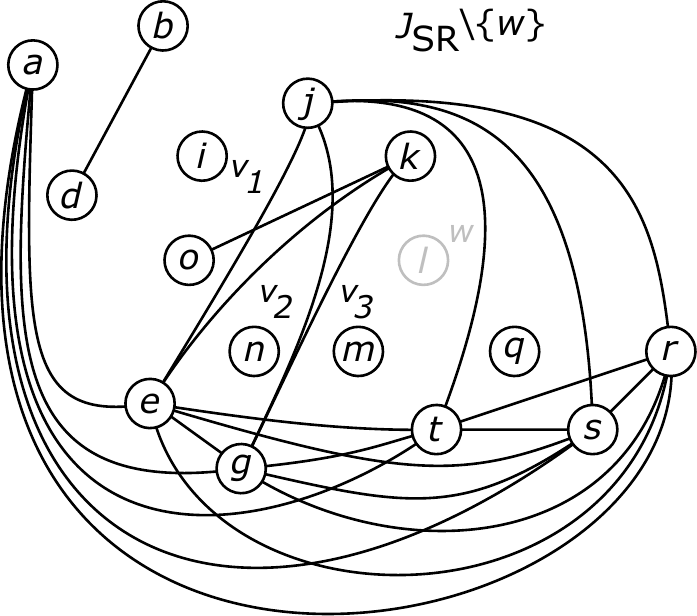}
\caption{The strong resolving graph $\SR{J}$ without vertex $w$, where $G_1$, $G_2$, $G_3$, $G_4$, and $H$ are from \Cref{figure2}. The four vertex sets $\{\capfont{a}\}$, $\{\capfont{e}\}$, $\{\capfont{g} \}$, $\{\capfont{r,s,t}\}$ form a complete 4-partite subgraph of $\SR{J}$. Vertex $\capfont{a}$ is the only vertex that is maximally distant from $u_1$ in $G_1$, vertex $\capfont{e}$ is the only vertex that is maximally distant from $u_2$ in $G_2$, vertex $\capfont{g}$ is the only vertex that is maximally distant from $u_3$ in $G_3$, and the vertices $\capfont{r,s,t}$ are the only vertices that are maximally distant from $u_4$ in $G_4$.
Also, the vertices $\capfont{l}$ and $\capfont{m}$ are maximally distant from $v_1$, the vertices $\capfont{j}$ and $\capfont{k}$ are maximally distant from $v_2$, and the vertices $\capfont{i}$ and $\capfont{j}$ are maximally distant from $v_3$ in $H$. Therefore, the vertex sets $\{\capfont{e}\}$ and $\{\capfont{j,k}\}$ form a complete bipartite subgraph in $\SR{J}$, the vertex sets $\{\capfont{g}\}$ and $\{\capfont{j,k}\}$ form a second complete bipartite subgraph and the vertex sets $\{\capfont{r,s,t}\}$ and $\{\capfont{j}\}$ form a third one. Since the vertices $\capfont{i,m}$ and $\capfont{n}$ are separation vertices in $J$ and the vertex $\capfont{l}$ will be a separation vertex later on (see \Cref{lemma8} and \Cref{lemma9}), they have no incident edges in $\SR{J}$.}
\label{figure4}
\end{figure}

The next two lemmas consider the case that graph $H$ has an additional vertex $w$ that can be used to attach $J$ to further graphs, see \Cref{figure4}.

%%%%%%%%%%%%%%%%%%%%%%%%%%%%%%%%%%%%%%%%%%%%%%%%%%%%%%%%%%%%
% Figure 4
%%%%%%%%%%%%%%%%%%%%%%%%%%%%%%%%%%%%%%%%%%%%%%%%%%%%%%%%%%%%

%%%%%%%%%%%%%%%%%%%%%%%%%%%%%%%%%%%%%%%%%%%%%%%%%%%%%%%%%%%%
% Lemma 8
%%%%%%%%%%%%%%%%%%%%%%%%%%%%%%%%%%%%%%%%%%%%%%%%%%%%%%%%%%%%

\begin{lemma}
\label{lemma8}
Let \Cref{prerequisite1} be given and $w \in V_H$. Let
$$U_0 = \left(\begin{array}{ll}
     & \MVC{\SR{H} \setminus \{v_1,\ldots,v_k,w\}} \\
\cup & \bigcup_{1 \leq i \leq k} \XVC{\SR{(G_i)} \setminus \{u_i\}}{\MD{G_i}{u_i}}
\end{array} \right)
$$
and
$$U_j = \left(\begin{array}{ll}
     & \XVC{\SR{H} \setminus \{v_1,\ldots,v_k,w\}}{\MD{H}{v_j}\setminus\{v_1,\ldots,v_k,w\}} \\
\cup & \MVC{\SR{(G_j)} \setminus \{u_j\}} \\
\cup & \bigcup_{1 \leq i \leq k, i\not=j} \XVC{\SR{(G_i)} \setminus \{u_i\}}{\MD{G_i}{u_i}}
\end{array} \right)
$$
for $j=1,\ldots,k$.

Each vertex sets $U_0,U_1,\ldots,U_k$ is a vertex cover for $\SR{J} \setminus \{w\}$, where at least one of them is a minimum vertex cover $\SR{J}$.

\begin{proof}
The only difference between \Cref{lemma7} and \Cref{lemma8} is the additional vertex $w$ in $H$, which is removed from the computations of the vertex covers, since it becomes a separation vertex in all further compositions. The correctness follows from the reasoning applied in \Cref{lemma7}.
\end{proof}
\end{lemma}

%%%%%%%%%%%%%%%%%%%%%%%%%%%%%%%%%%%%%%%%%%%%%%%%%%%%%%%%%%%%
% Lemma 9
%%%%%%%%%%%%%%%%%%%%%%%%%%%%%%%%%%%%%%%%%%%%%%%%%%%%%%%%%%%%

\begin{lemma}
\label{lemma9}
Let \Cref{prerequisite1} be given and $w \in V_H$. Then
$$\left(\begin{array}{ll}
     & \XVC{\SR{H} \setminus \{v_1,\ldots,v_k,w\}}{\MD{H}{w}\setminus\{v_1,\ldots,v_k,w\}} \\
\cup & \bigcup_{1 \leq i \leq k} \XVC{\SR{(G_i)} \setminus \{u_i\}}{\MD{G_i}{u_i}}.
\end{array} \right)
$$
is a minimum vertex cover for $\SR{J} \setminus \{w\}$ that contains all vertices of $\MD{J}{w}$.

\begin{proof}
A vertex cover for $\SR{G}$ that additionally contains all vertices from the sets $\MD{G_i}{u_i}$, $1 \leq i \leq k$, already covers all edges of the k-partite graph and the $k$ bipartite graphs as defined in the proof of \Cref{lemma7}. Therefore, no further case distinctions are necessary for the computation of $\XVC{\SR{J} \setminus \{w\}}{\MD{J}{w}}$, based on $\Cref{lemma2}$.
\end{proof}
\end{lemma}

%%%%%%%%%%%%%%%%%%%%%%%%%%%%%%%%%%%%%%%%%%%%%%%%%%%%%%%%%%%%
% Theorem 2
%%%%%%%%%%%%%%%%%%%%%%%%%%%%%%%%%%%%%%%%%%%%%%%%%%%%%%%%%%%%

The results from \Cref{lemma7,lemma8,lemma9} are summarized by the following theorem.

\begin{theorem}
\label{theorem2}
Let \Cref{prerequisite1} be given and $w \in V_H$. Then $$\MVC{\SR{J}},$$ $$\MVC{\SR{J}\setminus\{w\}},$$ and $$\XVC{\SR{J}\setminus\{w\}}{\MD{J}{w}}$$ are computable from $G_1,\ldots,G_k,H$, $u_1,\ldots,u_k$, $v_1,\ldots,v_k$, $w$, and the following vertex sets.
\begin{enumerate}
\item $\MVC{\SR{(G_i)}\setminus\{u_i\}}$, for $i=1,\ldots,k$, \newline
(used by \Cref{lemma7}: $U_j$, \Cref{lemma8}: $U_j$,),
\item $\XVC{\SR{(G_i)}\setminus\{u_i\}}{\MD{G_i}{\{u_i\}}}$ for $i=1,\ldots,k$, \newline
(used by \Cref{lemma7}: $U_0$, $U_j$, \Cref{lemma8}: $U_0$, $U_j$, \Cref{lemma9}),
\item
\begin{enumerate}
\item $\MVC{\SR{H}\setminus\{v_1,\ldots,v_k\}}$, \newline
(used by \Cref{lemma7}: $U_0$),
\item $\MVC{\SR{H}\setminus\{v_1,\ldots,v_k,w\}}$, \newline
(used by \Cref{lemma8}: $U_0$),
\end{enumerate}
\item
\begin{enumerate}
\item $\XVC{\SR{H}\setminus\{v_1,\ldots,v_k\}}{\MD{H}{\{v_i\}}\setminus\{v_1,\ldots,v_k\}}$ for $i=1,\ldots,k$, \newline
(used by \Cref{lemma7}: $U_j$),
\item $\XVC{\SR{H}\setminus\{v_1,\ldots,v_k,w\}}{\MD{H}{\{v_i\}}\setminus\{v_1,\ldots,v_k,w\}}$ for $i=1,\ldots,k$, \newline
(used by \Cref{lemma8}: $U_j$), and
\end{enumerate}
\item $\XVC{\SR{H}\setminus\{v_1,\ldots,v_k,w\}}{\MD{H}{\{w\}}\setminus\{v_1,\ldots,v_k,w\}}$, \newline
(used by \Cref{lemma9}).
\end{enumerate}
\end{theorem}

In the next section we show how a minimum strong resolving set for a graph $G$ can be efficiently computed using \Cref{lemma7,lemma8,lemma9}, provided that a minimum strong resolving set for the biconnected components can be efficiently computed.

%%%%%%%%%%%%%%%%%%%%%%%%%%%%%%%%%%%%%%%%%%%%%%%%%%%%%%%%%%%%
% Section 6
%%%%%%%%%%%%%%%%%%%%%%%%%%%%%%%%%%%%%%%%%%%%%%%%%%%%%%%%%%%%

\section{The algorithmic frame}
\label{section6}

\begin{figure}[H]
\center
\includegraphics[width=360pt]{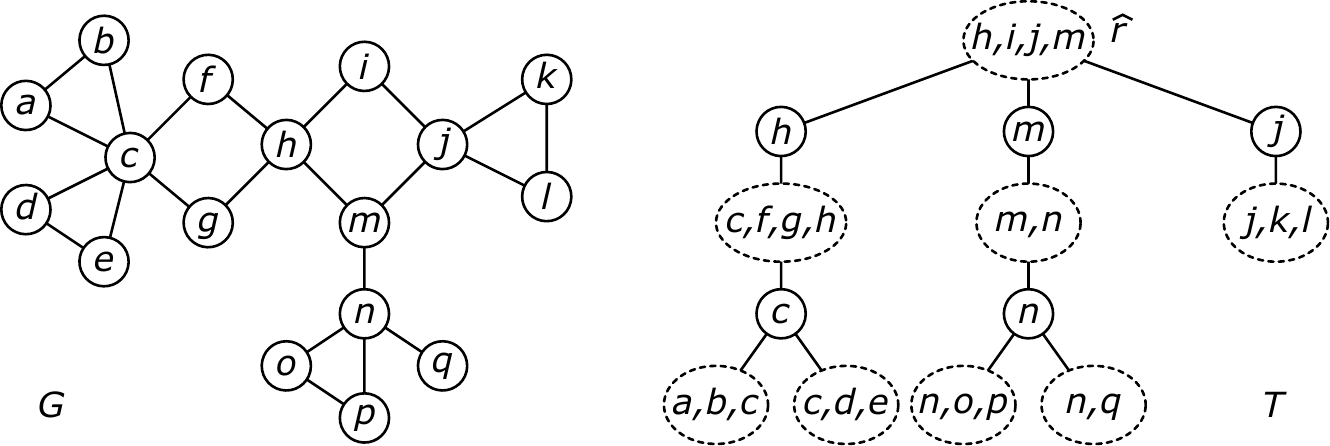}
\caption{A graph $G$ to the left and its decomposition tree $T$ to the right. The dashed circles are the b-nodes of $T$ for the biconnected components of $G$. The other nodes are the s-nodes of $T$ for the separation vertices in $G$. The b-node for the biconnected component of $G$ induced by the vertices $\capfont{h,i,j,m}$ has been selected as the root $\hat{r}$ of $T$.}
\label{figure5}
\end{figure}

The given graph $G$ is first decomposed into its biconnected components. Edges, whose end vertices are separation vertices or vertices of degree one, are also regarded as biconnected components. Then the {\em decomposition tree $T$} for $G$ is built. We use variable names with a hat symbol for nodes in trees to distinguish them from the vertices in graphs. Tree $T$ contains a so-called {\em b-node} for each biconnected component of $G$ and a so-called {\em s-node} for each separation vertex of $G$. A b-node $\hat{u}$ and an s-node $\hat{v}$ are connected by an edge $\{\hat{u},\hat{v}\}$ in $T$ if and only if the separation vertex for $\hat{v}$ is part of the biconnected component for $\hat{u}$. The preprocessing to compute $T$ can be done in linear time. \Cref{figure5} shows an example of such a decomposition.

%%%%%%%%%%%%%%%%%%%%%%%%%%%%%%%%%%%%%%%%%%%%%%%%%%%%%%%%%%%%
% Figure 5
%%%%%%%%%%%%%%%%%%%%%%%%%%%%%%%%%%%%%%%%%%%%%%%%%%%%%%%%%%%%

Computing a minimum strong resolving set for $G$, or equivalently, computing a minimum vertex cover for $\SR{G}$, can now be done via a bottom-up processing of $G$ according to the decomposition tree $T$. Tree $T$ is first oriented by choosing any b-node $\hat{r}$ of $T$ as the root of $T$. All leaves of $T$ are b-nodes. The predecessor nodes of the leaves are s-nodes for separation vertices of $G$ that connect the biconnected components of the leafs to the rest of $G$. The predecessor nodes of the predecessor nodes of the leaves of $T$ are again b-nodes for biconnected components of $G$ at which the biconnected components for the leaves are linked via the separation vertices, and so on.

For a b-node $\hat{u}$ of $T$ let $G(\hat{u})$ be the biconnected component for $\hat{u}$ and $\widetilde{G}(\hat{u})$ be the subgraph of $G$ induced by the vertices of all biconnected components of the b-nodes in the subtree of $T$ with root $\hat{u}$. For an s-node $\hat{v}$ let $\sv{\hat{v}}$ be the separation vertex for $\hat{v}$.

%%%%%
A minimum strong resolving set for $G$ is equal to $\MVC{\SR{\widetilde{G}(\hat{r})}}$, which can be calculated based on informations about $G(\hat{r})$ and the subgraphs attached to it, using \Cref{lemma7,lemma8,lemma9}.

\begin{figure}[H]
\center
\includegraphics[width=380pt]{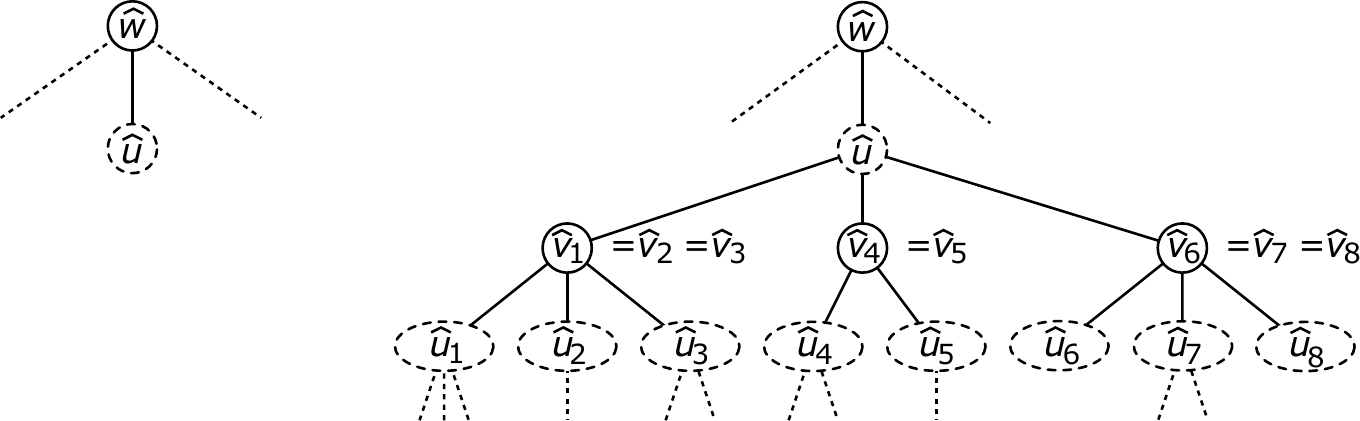}
\caption{The left-hand side shows the first case where the b-node $\hat{u}$ is a leaf of $T$. The right-hand side shows the second case in which the b-node $\hat{u}$ is an inner node of $T$.}
\label{figure6}
\end{figure}

To describe the vertex sets we calculate more detailed, we consider the following cases. Let $\hat{u}$ be a b-node of $T$.
\begin{enumerate}
\item
If $\hat{u}$ is a leaf with predecessor s-node $\hat{w}$, then we calculate the two sets
$$\MVC{\SR{\widetilde{G}(\hat{u})}\setminus\{\sv{\hat{w}}\}} \quad \text{and} \quad \XVC{\SR{\widetilde{G}(\hat{u})}\setminus\{\sv{\hat{w}}\}}{\MD{\widetilde{G}(\hat{u})}{\sv{\hat{w}}}}$$
directly from the biconnected subgraph $\widetilde{G}(\hat{u}) = G(\hat{u})$ of $G$. \Cref{figure6} shows an example of this case on the left-hand side. If $\hat{u}$ is a leaf without a predecessor s-node, then $G(\hat{u}) = G$ and we calculate $\MVC{\SR{G(\hat{u})}}$ directly.
\item
If $\hat{u}$ is an inner b-node of $T$ with predecessor s-node $\hat{w}$, let $\hat{u}_1, \cdots, \hat{u}_k$ be the successor b-nodes of the successor s-nodes of $\hat{u}$, and $\hat{v}_1, \ldots,\hat{v}_k$ be the predecessor s-nodes of $\hat{u}_1, \cdots, \hat{u}_k$. Note that $\hat{v}_1, \ldots,\hat{v}_k$ do not need to be distinct. \Cref{figure6} shows an example of this case on the right-hand side. If we replace the separation vertex $\sv{\hat{v}_i}$ in subgraph $\widetilde{G}(\hat{u}_i)$ by a new vertex $w_i$ such that all $k$ graphs $\widetilde{G}(\hat{u}_1),\ldots,\widetilde{G}(\hat{u}_k)$, and the biconnected component  $G(\hat{u})$ are vertex-disjoint, then $$\widetilde{G}(\hat{u}) = (\widetilde{G}(\hat{u}_1) \cup \ldots \cup \widetilde{G}(\hat{u}_k)) \MERGE{w_1,\ldots,w_k}{\sv{\hat{v}_1},\ldots,\sv{\hat{v}_k}} G(\hat{u}).$$
We then calculate the two sets $$\MVC{\SR{\widetilde{G}(\hat{u})}\setminus\{\sv{\hat{w}}\}} \quad \text{and} \quad \XVC{\SR{\widetilde{G}(\hat{u})}\setminus\{\sv{\hat{w}}\}}{\MD{\widetilde{G}(\hat{u})}{\sv{\hat{w}}}}$$ using \Cref{lemma8,lemma9}. Those lemmas require the vertex sets\\ $\MVC{\SR{G(\hat{u})}\setminus \{\sv{\hat{v}_1},\ldots,\sv{\hat{v}_k},\sv{\hat{w}}\}}$,\\ $\XVC{\SR{G(\hat{u})} \setminus \{\sv{\hat{v}_1},\ldots,\sv{\hat{v}_k},\sv{\hat{w}}\}}{\MD{G(\hat{u})}{\sv{\hat{v}_i}}\setminus \{\sv{\hat{v}_1},\ldots,\sv{\hat{v}_k},\sv{\hat{w}}\}}$, and\\ $\XVC{\SR{G(\hat{u})} \setminus \{\sv{\hat{v}_1},\ldots,\sv{\hat{v}_k},\sv{\hat{w}}\}}{\MD{G(\hat{u})}{\sv{\hat{w}}}\setminus \{\sv{\hat{v}_1},\ldots,\sv{\hat{v}_k},\sv{\hat{w}}\}}$.
\item
If $\hat{u} = \hat{r}$, then we calculate the set $$\MVC{\SR{\widetilde{G}(\hat{u})}}$$ using \Cref{lemma7}. This lemma requires the vertex sets\\ $\MVC{\SR{G(\hat{u})}\setminus \{\sv{\hat{v}_1},\ldots,\sv{\hat{v}_k}\}}$ and\\ $\XVC{\SR{G(\hat{u})} \setminus \{\sv{\hat{v}_1},\ldots,\sv{\hat{v}_k}\}}{\MD{G(\hat{u})}{\sv{\hat{v}_i}}\setminus \{\sv{\hat{v}_1},\ldots,\sv{\hat{v}_k}\}}$.
\end{enumerate}

%%%%%%%%%%%%%%%%%%%%%%%%%%%%%%%%%%%%%%%%%%%%%%%%%%%%%%%%%%%%
% Figure 6
%%%%%%%%%%%%%%%%%%%%%%%%%%%%%%%%%%%%%%%%%%%%%%%%%%%%%%%%%%%%

Recall that \Cref{lemma7,lemma8} focus on the $k$ bipartite subgraphs of $\SR{\widetilde{G}(\hat{u})}$ (formerly $\SR{J}$), between vertices of $\widetilde{G}(\hat{u}_i)$ (formerly $G_i$) and vertices of $G(\hat{u})$ (formerly $H$). To determine which one of those $k$ bipartite subgraphs, if any, needs to be covered by vertices of $G(\hat{u})$, we defined the sets $U_j$ for $0\leq j\leq k$. The sizes of those sets can be calculated as follow. Note that we focus on \Cref{lemma8}, since \Cref{lemma7} is only needed for the root $\hat{r}$, with the only difference being the separation vertex $s(\hat{w})$. Let
$$h_i = \vert\XVC{\SR{G(\hat{u})} \setminus \{\sv{\hat{v}_1},\ldots,\sv{\hat{v}_k},\sv{\hat{w}}\}}{\MD{G(\hat{u})}{\sv{\hat{v}_i}}\setminus \{\sv{\hat{v}_1},\ldots,\sv{\hat{v}_k},\sv{\hat{w}}\}}\vert$$ and let $$h'_i = \vert\XVC{\SR{\widetilde{G}(\hat{u}_i)} \setminus\{\sv{\hat{v}_i}\}}{\MD{\widetilde{G}(\hat{u}_i)}{\sv{\hat{v}_i}}}\vert.$$ Then $h_i$ is the number of vertices needed to cover the $i$-th bipartite subgraph with vertices of $G(\hat{u})$ and $h'_i$ is the number of vertices needed to cover it with vertices of $\widetilde{G}(\hat{u}_i)$. Therefore, $\vert U_i\vert = \vert U_0\vert + h_i - h'_i$ for $1\leq i\leq k$.

The previous considerations mainly come down to the following. To compute the strong metric dimension of a graph based on the biconnected components, the size of\\ $\XVC{\SR{H} \setminus W}{\MD{H}{u} \setminus W}$ for a given graph $H$, a vertex $u$ of $H$, and a vertex set $W$ of $H$ has to be determined. If this size can be calculated in total linear (respectively polynomial) time for $k$ different vertices $u$ for an integer $i$ and a (possibly empty) vertex set $W$ for every biconnected component $H$, it is generally possible to compute the strong metric dimension of the whole graph in total linear (respectively polynomial) time as well. Therefore the total running time of the entire computation depends on the computations of the minimum vertex covers for the induced subgraphs $\SR{G(\hat{u})}$ of the b-nodes $\hat{u}$ of $T$, as well as the modified  vertex covers including the vertices, which are maximally distant from the separation vertices. In the next section we explain the processing time in more detail using three examples. 

%%%%%%%%%%%%%%%%%%%%%%%%%%%%%%%%%%%%%%%%%%%%%%%%%%%%%%%%%%%%
% Section 7
%%%%%%%%%%%%%%%%%%%%%%%%%%%%%%%%%%%%%%%%%%%%%%%%%%%%%%%%%%%%

\section{Three examples}
\label{section7}

%%%%%%%%%%%%%%%%%%%%%%%%%%%%%%%%%%%%%%%%%%%%%%%%%%%%%%%%%%%%

Let again $\hat{u}$, $\hat{u}_1, \ldots, \hat{u}_k$ be b-nodes of $T$ and $\hat{w}$, $\hat{v}_1, \ldots, \hat{v}_k$ be the predecessor s-nodes of them in the manner they are used in \Cref{section6}, see also \Cref{figure6} on the right. The following examples show how to compute
\begin{itemize}
\item
$\MVC{\SR{\widetilde{G}(\hat{u})}}$,
\item
$\MVC{\SR{\widetilde{G}(\hat{u})}\setminus\{\sv{\hat{w}}\}}$, and 
\item
$\XVC{\SR{\widetilde{G}(\hat{u})}\setminus\{\sv{\hat{w}}\}}{\MD{\widetilde{G}(\hat{u})}{\sv{\hat{w}}}}$
\end{itemize}
in linear time if the biconnected components have a specific structure. % if
%\begin{itemize}
%\item $G(\hat{u})$,
%\item $\MVC{\SR{\widetilde{G}(\hat{u}_i)} \setminus\{\sv{\hat{v}_i}\}}$, and
%\item $\XVC{\SR{\widetilde{G}(\hat{u}_i)} \setminus\{\sv{\hat{v}_i}\}}{\MD{\widetilde{G}(\hat{u}_i)}{\sv{\hat{v}_i}}}$
%for $1 \leq i \leq k$
%\end{itemize}
%are given, see \Cref{lemma7,lemma8,lemma9}.
In order to compute the sets above in linear time, we show
\begin{enumerate}
\item how to compute 
$$\XVC{\SR{G(\hat{u}_i)} \setminus \{\sv{\hat{v}_1},\ldots,\sv{\hat{v}_k},\sv{\hat{w}}\}}{\MD{G(\hat{u}_i)}{\sv{\hat{w}}}\setminus\{\sv{\hat{v}_1},\ldots,\sv{\hat{v}_k},\sv{\hat{w}}\}}$$
in linear time and
\item how to decide which of the sets $U_0,U_1,\ldots,U_k$ from \Cref{lemma8,lemma9} is the smallest set in total linear time.
\end{enumerate}

%%%%%%%%%%%%%%%%%%%%%%%%%%%%%%%%%%%%%%%%%%%%%%%%%%%%%%%%%%%%
% Grids
%%%%%%%%%%%%%%%%%%%%%%%%%%%%%%%%%%%%%%%%%%%%%%%%%%%%%%%%%%%%

\subsection{Grids}
\label{section_grids}

As our first example we consider grids. Let $G$ be an $n\times m$ grid with vertex set $$\{x_{1,1},\ldots,x_{1,m},\ldots,x_{n,1},\ldots,x_{n,m}\},$$ with $n,m\geq 2$. The strong resolving graph of $G$ only contains the two edges $\{x_{1,1},x_{n,m}\}$ and $\{x_{1,m},x_{n,1}\}$. Therefore, computing a vertex cover for $\SR{G}$ is fairly simple and $\MD{G}{x_{i,j}}$ can be determined easily as well.

\begin{observation}
Let $G$ be an $n\times m$ grid as defined above and let $x_{i,j}$ with $1\leq i\leq n$ and $1\leq j\leq m$ be a vertex of $G$.
\begin{enumerate}
\item
If $i\in\{1,n\}$ and $j\in\{1,m\}$, then $\MD{G}{x_{i,j}} = \{x_{n+1-i,m+1-j}\}$. That is, if $x_{i,j}$ is a corner vertex of the grid, then $\MD{G}{x_{i,j}}$ only contains the opposite corner vertex.
\item
If $i\in\{1,n\}$ and $j\notin\{1,m\}$, then $\MD{G}{x_{i,j}} = \{x_{n+1-i,1}, x_{n+1-i,m}\}$. Analogously, if $i\notin\{1,n\}$ and $j\in\{1,m\}$, then $\MD{G}{x_{i,j}} = \{x_{1,m+1-j}, x_{n,m+1-j}\}$. That is, if $x_{i,j}$ is an edge vertex of the grid, then $\MD{G}{x_{i,j}}$ contains the two corner vertices on the opposite edge.
\item
If $i\notin\{1,n\}$ and $j\notin\{1,m\}$, then $\MD{G}{x_{i,j}} = \{x_{1,1}, x_{n,1}, x_{1,m}, x_{n,m}\}$. That is, if $x_{i,j}$ is an inner vertex of the grid, then $\MD{G}{x_{i,j}}$ contains all four corner vertices.
\end{enumerate}
\end{observation}

%%%%%%%%%%%%%%%%%%%%%%%%%%%%%%%%%%%%%%%%%%%%%%%%%%%%%%%%%%%%
% Figure 13
%%%%%%%%%%%%%%%%%%%%%%%%%%%%%%%%%%%%%%%%%%%%%%%%%%%%%%%%%%%%
\begin{figure}[H]
\center
\includegraphics[width=380pt]{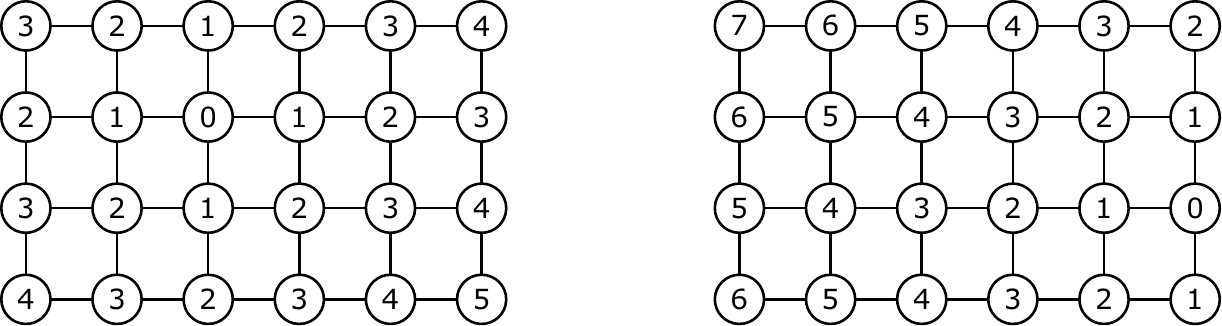}
\caption{A grid $G_{4,6}$ and the distance of each vertex from the inner vertex $x_{2,3}$ on the left, and from the edge vertex $x_{3,6}$ on the right. This example illustrates the distribution of distances in a grid, that only corner vertices are maximally distant from other vertices, and which corner vertices those are for different types of vertex in the grid.}
\label{figure13}
\end{figure}

\Cref{figure13} shows an example of a $4\times 6$ grid and the distances of the vertices to illustrate the previous observation. 
Since $\SR{G}$ only contains two edges as mentioned above and with the previous observation, it is straight forward to see, that the size of $\XVC{\SR{G} \setminus W}{\MD{G}{x_{i,j}} \setminus W}$ can be calculated in constant time for every vertex $x_{i,j}$ and an arbitrary vertex set $W$ if the position of each vertex inside the grid is known. Therefore, the following theorem follows.

\begin{theorem}
\label{theorem3}
A minimum strong resolving set for a graph $G=(V,E)$, in that each biconnected component is a grid, can be computed in time ${\mathcal O}(|V| + |E|)$.
\end{theorem}

%%%%%%%%%%%%%%%%%%%%%%%%%%%%%%%%%%%%%%%%%%%%%%%%%%%%%%%%%%%%
% Cycles
%%%%%%%%%%%%%%%%%%%%%%%%%%%%%%%%%%%%%%%%%%%%%%%%%%%%%%%%%%%%

\subsection{Cycles}
\label{section_cycles}

\begin{figure}[ht]
\center
\includegraphics[width=\textwidth]{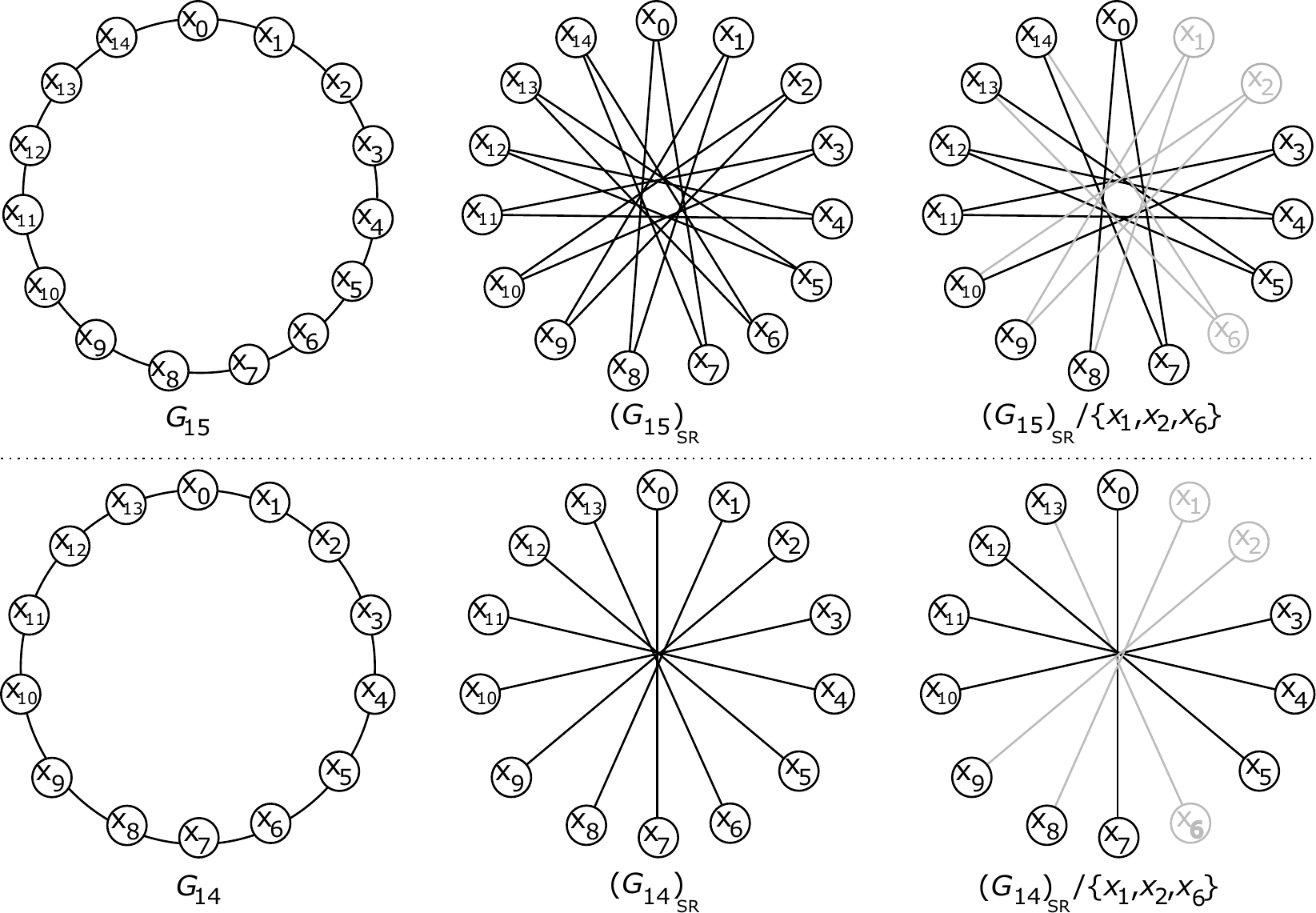}
\caption{Two circles of 15 and 14 vertices on the left, the corresponding strong resolving graphs $\SR{(G_{15})}$ and $\SR{(G_{14})}$ in the middle, and the strong resolving graphs without the vertices $x_1,x_2,x_6$ on the right.}
\label{figure12}
\end{figure}

Consider the example in which each biconnected component $G(\hat{u})$ of $G$ is a cycle with at least three vertices $x_0,x_1,\ldots,x_{n-1}$, $n \geq 3$. The edge set of $G(\hat{u})$ is $$\{\{x_0,x_1\},\{x_1,x_2\},\ldots,\{x_{n-1},x_0\}\}.$$
If $n$ is even, then $\SR{G(\hat{u})}$ consists of $n/2$ edges, such that no two edges have a vertex in common. In this case, $\MD{G(\hat{u})}{x_i}$ contains the single vertex $x_j$ with $j=(i+n/2) \bmod n$. If $n$ is odd, then $\SR{G(\hat{u})}$ is a cycle and $\MD{G(\hat{u})}{x_i}$ contains the two vertices $x_j$ and $x_{j+1}$ with $j=(i+\lfloor n/2 \rfloor) \bmod n$, see \Cref{figure12} for an example.

In both cases, $\SR{G(\hat{u})} \setminus \{\sv{\hat{v}_1},\ldots,\sv{\hat{v}_k},\sv{\hat{w}}\}$ is a collection of paths. A minimum vertex cover of
$$\SR{G(\hat{u})} \setminus \{\sv{\hat{v}_1},\ldots,\sv{\hat{v}_k},\sv{\hat{w}}\}$$ has $\lfloor l/2 \rfloor$ vertices from each path of those paths with $l$ vertices.
However,
$$\XVC{\SR{G(\hat{u})} \setminus \{\sv{\hat{v}_1},\ldots,\sv{\hat{v}_k},\sv{\hat{w}}\}}{\MD{G(\hat{u})}{x_i}\setminus \{\sv{\hat{v}_1},\ldots,\sv{\hat{v}_k},\sv{\hat{w}}\}}$$
may have some additional vertices depending on which paths the vertices from $\MD{G(\hat{u})}{x_i}$ belong to.  
Since $\MD{G(\hat{u})}{x_i}$ contains at most two vertices, the size of
$$\XVC{\SR{G(\hat{u})} \setminus \{\sv{\hat{v}_1},\ldots,\sv{\hat{v}_k},\sv{\hat{w}}\}}{\MD{G(\hat{u})}{x_i}\setminus \{\sv{\hat{v}_1},\ldots,\sv{\hat{v}_k},\sv{\hat{w}}\}}$$
can easily be computed in constant time. Here the length of the paths on which the vertices are located, and in addition, if both vertices are located on the same path $p$, the distance between them on $p$ must be taken into account. Also observe, that the vertices of $\MD{G(\hat{u})}{x_i}$ are always at the end of paths if $x_i$ is one of the vertices which are removed from $\SR{G(\hat{u})}$.

%%%%%%%%%%%%%%%%%%%%%%%%%%%%%%%%%%%%%%%%%%%%%%%%%%%%%%%%%%%%
% Figure 12
%%%%%%%%%%%%%%%%%%%%%%%%%%%%%%%%%%%%%%%%%%%%%%%%%%%%%%%%%%%%

Overall the size of $\XVC{\SR{H} \setminus W}{\MD{H}{u} \setminus W}$ for $k$ different vertices $u$ and a vertex set $W$ can be computed in total linear time if $H$ is a cycle. Thus, all calculations required to determine the strong metric dimension of $G$ can be done in total linear time if all biconnected components of $G$ are cycles, using the algorithm from \Cref{section6}. Also, which of the sets $U_0,U_1,\ldots,U_k$ is the smallest can be determined using the values $h_i$ and $h'_i$ mentioned at the end of \Cref{section6}.

%%%%%%%%%%%%%%%%%%%%%%%%%%%%%%%%%%%%%%%%%%%%%%%%%%%%%%%%%%%%
% Theorem 3
%%%%%%%%%%%%%%%%%%%%%%%%%%%%%%%%%%%%%%%%%%%%%%%%%%%%%%%%%%%%

\begin{theorem}
\label{theorem3}
A minimum strong resolving set for a graph $G=(V,E)$, in that each biconnected component is a cycle, can be computed in time ${\mathcal O}(|V| + |E|)$.
\end{theorem}

\subsection{Co-Graphs}
A more complex example arises if each biconnected component $G(\hat{u})$ of $G$ is a co-graph. Co-graphs can be defined as follows.

%%%%%%%%%%%%%%%%%%%%%%%%%%%%%%%%%%%%%%%%%%%%%%%%%%%%%%%%%%%%
% Definition 4
%%%%%%%%%%%%%%%%%%%%%%%%%%%%%%%%%%%%%%%%%%%%%%%%%%%%%%%%%%%%

\begin{definition}[\bf Co-Graphs and Co-Trees]
\label{definition4}
\cite{CLB81}
\begin{itemize}
\item A graph $G$ that consists of a single vertex $u$ is a co-graph. The co-tree $T$ for $G$ consists of a single node $\hat{u}$ associated with vertex $u$ of $G$. Node $\hat{u}$ is the {\em root} of $T$. Let $\vertex{\hat{u}} = u$ and $\node{u} = \hat{u}$. Note that $\vertex{\hat{u}}$ is only defined for leaves $\hat{u}$ of $T$.

\item If $G_1=(V_1,E_1)$ and $G_2=(V_2,E_2)$ are two co-graphs, then the disjoint {\em union} of $G_1$ and $G_2$, denoted by $G_1 \union G_2$, is a co-graph $G$ with vertex set $V_1\cup V_2$ and edge set $E_1\cup E_2$. Let $T_1$ and $T_2$ be co-trees for $G_1$ and $G_2$ with root $\hat{u}_1$ and $\hat{u}_2$, respectively. Then tree $T$ defined by the disjoint union of $T_1$ and $T_2$ with an additional node $\hat{u}$ and two additional edges $\{\hat{u},\hat{u}_1\}$ and $\{\hat{u},\hat{u}_2\}$ is a co-tree for $G$. Node $\hat{u}$ is the {\em root} of $T$ labelled by $\union$. Node $\hat{u}_1$ and $\hat{u}_2$ are successor nodes of $\hat{u}$. Node $\hat{u}$ is the predecessor node of $\hat{u}_1$ and $\hat{u}_2$.

\item If $G_1=(V_1,E_1)$ and $G_2=(V_2,E_2)$ are two co-graphs, then the {\em join} of $G_1$ and $G_2$, denoted by $G_1 \join G_2$, is a co-graph with vertex set $V_1\cup V_2$ and edge set $E_1\cup E_2\cup \{\{u,v\}\ \vert\ u\in V_1, v\in V_2\}$. Let $T_1$ and $T_2$ be co-trees for $G_1$ and $G_2$ with root $\hat{u}_1$ and $\hat{u}_2$, respectively. Then tree $T$ defined by the disjoint union of $T_1$ and $T_2$ with an additional node $\hat{u}$ and two additional edges $\{\hat{u},\hat{u}_1\}$ and $\{\hat{u},\hat{u}_2\}$ is a co-tree for $G$. Node $\hat{u}$ is the {\em root} of $T$ labelled by $\join$. Node $\hat{u}_1$ and $\hat{u}_2$ are successor nodes of $\hat{u}$. Node $\hat{u}$ is the predecessor node of $\hat{u}_1$ and $\hat{u}_2$.

\end{itemize}
\end{definition}

Co-graphs can be recognized in linear time, see \cite{JO95}. This includes the computation of a co-tree. Co-graphs are the graphs that do not contain an induced $P_4$, a path with four vertices, or in other words, connected co-graphs are graphs with diameter at most $2$. 

Two adjacent vertices $u$ and $v$ of a graph $G$ are {\em true twins} if $N_G[u] = N_G[v]$. In \cite{SW21} it is shown that the strong resolving graph of a connected co-graph is again a co-graph and that two vertices $u$ and $v$ are mutually maximally distant in $G$ if and only if they are either not adjacent in $G$ or they are true twins in $G$.

\begin{figure}[ht]
\center
\includegraphics[width=240pt]{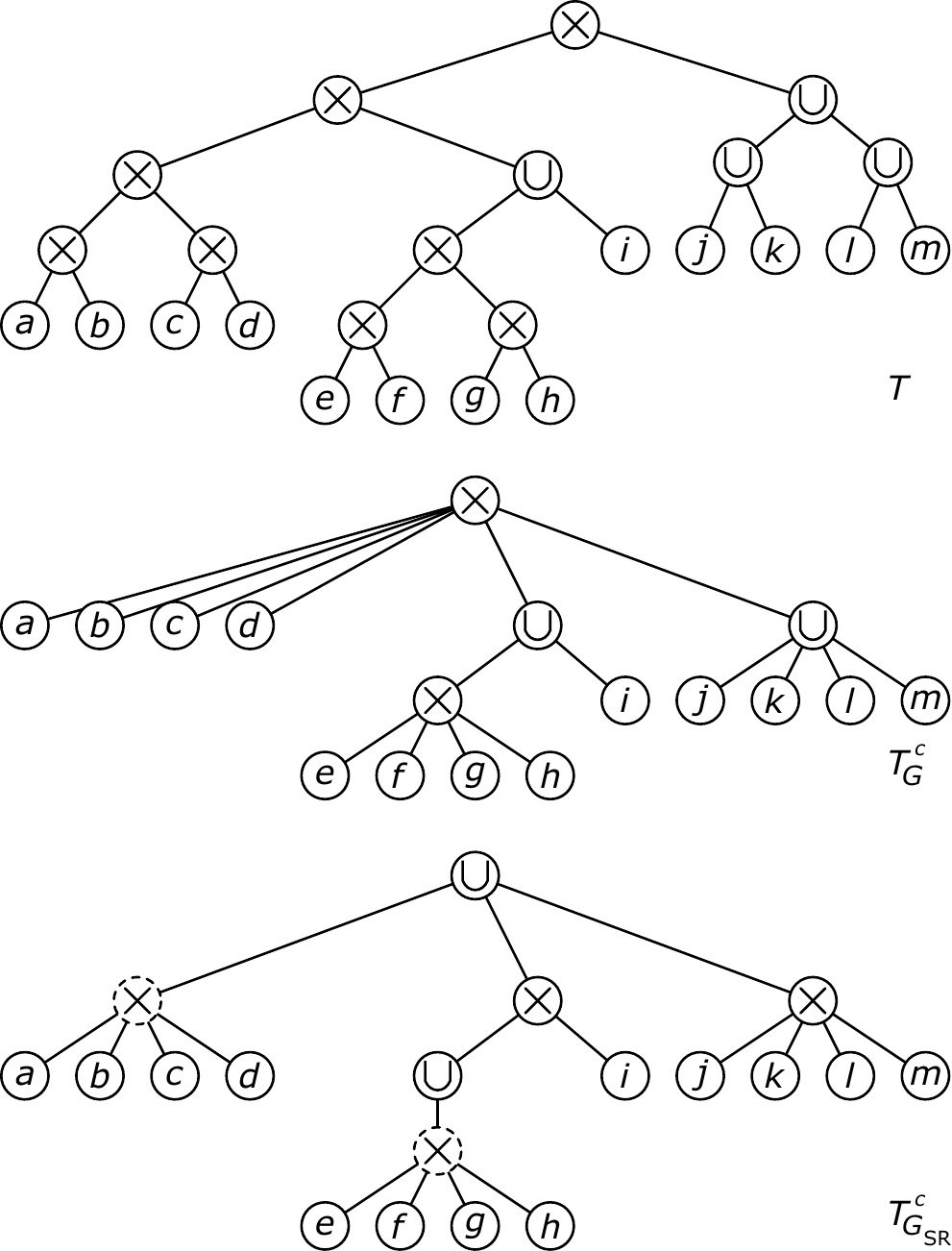}
\caption{A co-tree $T$ for a co-graph $G$, a canonical co-tree $T^{\text c}_G$ for $G$, and a canonical co-tree for $T^{\text c}_{\SR{G}}$ for $\SR{G}$. The two nodes with the dashed circles are twin-join nodes in $T^{\text c}_{\SR{G}}$. These nodes are not in $T^c_G$.}
\label{figure7}
\end{figure}

For the analysis of co-graphs, the so-called {\em canonical co-tree} is a useful data structure. The canonical co-tree results from combining successive union and join operations into one union and join operation, see also \Cref{figure7}. It can also be computed in linear time, see \cite{HP05}. In a {\em canonical co-tree} each inner node may have more than two successor nodes. The successor nodes of a union node are join nodes or leafs, the successor nodes of a join node are union nodes or leafs. Two vertices $u$ and $v$ of $G$ are true twins in $G$ if and only if $\node{u}$ and $\node{v}$ are leaves of a common join node in the {\em canonical co-tree} for $G$.

A canonical co-tree $T^{\text c}_{\SR{G}}$ for $\SR{G}$ can be easily constructed from a canonical co-tree $T^{\text c}_G$ for $G$ by first transforming the union nodes into join nodes and the join nodes into union nodes. The corresponding graph of the new co-tree has an edge between two vertices $u$ and $v$ if and only if there was no edge between $u$ and $v$ beforehand. If now a union node $\hat{u}$ of  $T^{\text c}_{\SR{G}}$ (that is a join node in $T^{\text c}_G$) has two or more successor nodes $\hat{u}_1,\ldots,\hat{u}_k$ that are leaves, then $\hat{u}_1,\ldots,\hat{u}_k$ are true twins in $G$ and we detach them from $\hat{u}$ in $T^{\text c}_{\SR{G}}$, insert a new join node $\hat{w}$ in $T^{\text c}_{\SR{G}}$ as a successor node of $\hat{u}$ and append the detached leaves $\hat{u}_1,\ldots,\hat{u}_k$ to the new join node $\hat{w}$. We mark this new join node $\hat{w}$ as a {\em twin-join node}, see also \Cref{figure7}. If all successor nodes of $\hat{u}$ in $T^{\text c}_{\SR{G}}$ were leaves before, then $\hat{u}$ might have only the one successor node $\hat{w}$ after the modification. It is important for our forthcoming processing to preserve this structure and to not clean it up by attaching the leaves of $\hat{w}$ to the predecessor node of $\hat{u}$. The resulting tree $T^{\text c}_{\SR{G}}$ is a canonical co-tree for $\SR{G}$, because two vertices $u$ and $v$ are adjacent in $\SR{G}$ if and only if they are either not adjacent in $G$ or they are true twins in $G$ if and only if they are mutually maximally distant in $G$.

%%%%%%%%%%%%%%%%%%%%%%%%%%%%%%%%%%%%%%%%%%%%%%%%%%%%%%%%%%%%
% Figure 7
%%%%%%%%%%%%%%%%%%%%%%%%%%%%%%%%%%%%%%%%%%%%%%%%%%%%%%%%%%%%

The size of a minimum strong resolving set for a co-graph $G$ can be computed in linear time by computing the size of a minimum vertex cover for co-graph $\SR{G}$. We use the following notations to describe this well-known computation procedure. Let $T$ be a co-tree for a co-graph $G$ with root $\hat{r}$. For a node $\hat{u}$ of $T$, let $T(\hat{u})$ be the subtree of $T$ with root $\hat{u}$ and $\text{n}(\hat{u})$ be the number of leafs in $T(\hat{u})$.

\begin{figure}[ht]
\center
\includegraphics[width=240pt]{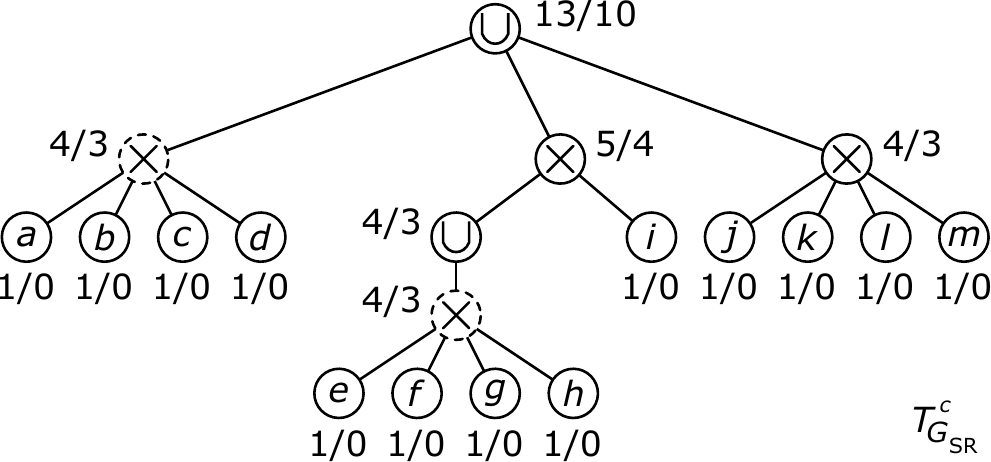}
\caption{The computation of a minimum vertex cover for co-graph $\SR{G}$ defined by the canonical co-tree $T^c_{\SR{G}}$ of \Cref{figure7}. The nodes $\hat{u}$ of tree $T^c_{\SR{G}}$ are labeled $\text{n}(\hat{u})/\text{vc}(\hat{u})$.}
\label{figure8}
\end{figure}

The size of a minimum vertex cover for a co-graph $\SR{G}$ with canonical co-tree $T^c_{\SR{G}}$ and root $\hat{r}$ can be computed in linear time by the bottom-up processing of $T$ with Algorithm\,1. The result is $\text{vc}(\hat{r})$, see also \Cref{figure8}.

\begin{algorithm}[ht]
\SetKwProg{Fn}{Algorithm\,1\,}{$(T^c_{\SR{G}})$}{}
\Fn{}{
	\For{$($each node $\hat{u}$ of $T^c_{\SR{G}})$}{
		\tcp{\it $-1$ means undefined}
		$\text{vc}(\hat{u}) \leftarrow -1$;
	}
	\For{$($each leaf $\hat{u}$ of $T^c_{\SR{G}})$}{
		$\text{vc}(\hat{u}) \leftarrow 0$;
	}
	\While{
	$\left(
	\begin{array}{l}\text{there is a node } \hat{u} \text{ of } T^c_{\SR{G}} \text{ with successor nodes } \hat{u}_1,\ldots,\hat{u}_k \\
	\text{such that } \text{\rm vc}({\hat{u}}) = -1 \text{ and } \text{\rm vc}({\hat{u}_i}) \geq 0, 1 \leq i \leq k
	\end{array}
	\right)$}{
		\ \\
%		Let $\hat{u}_1,\ldots,\hat{u}_k$ be the successor nodes of $\hat{u}$; \\
		\If{$(\hat{u}$ is a union node in $T^c_{\SR{G}})$}{
			$\text{vc}({\hat{u}}) \leftarrow \text{vc}(\hat{u}_1) + \cdots + \text{vc}(\hat{u}_k)$;
		}
		\Else
		{
			\tcp{\it $\hat{u}$ is a join node in $T^c_{\SR{G}}$}
			$\text{vc}({\hat{u}}) \leftarrow \displaystyle \min_{1 \leq i \leq k} (\text{n}(\hat{u}) - \text{n}(\hat{u}_i) + \text{vc}(\hat{u}_i))$; \\
		}
	}
}
\end{algorithm}

%%%%%%%%%%%%%%%%%%%%%%%%%%%%%%%%%%%%%%%%%%%%%%%%%%%%%%%%%%%%
% Figure 8
%%%%%%%%%%%%%%%%%%%%%%%%%%%%%%%%%%%%%%%%%%%%%%%%%%%%%%%%%%%%

However, we need to compute the size of a minimum vertex set that is a vertex cover for $\SR{G}$ which contains all vertices that are maximally distant from a vertex $u$ in $G$.

The vertices $v$ that are maximally distant from a vertex $u$ in co-graph $G$ can be specified as follows. Let $T^c_G$ be a canonical co-tree for $G$ and $\hat{w}$ be the first common ancestor of $\hat{u}= \node{u}$ and $\hat{v} = \node{v}$ in $T^c_G$.

\begin{enumerate}
\item
If $\hat{w}$ is a union node, then $u$ and $v$ are not adjacent in $G$ and thus $v$ is maximally distant from $u$ (and $u$ is maximally distant from $v$). Note that connected co-graphs have diameter at most 2.
\item
If $\hat{w}$ is a join node, then $u$ and $v$ are adjacent in $G$ and $v$ is maximally distant from $u$ if and only if $N_G[v] \subseteq N_G[u]$. This is the case if and only if all nodes on the path between $\hat{u}$ and $\hat{w}$ in $T$ are join nodes. If $T^c_G$ is a canonical co-tree, then this is the case if and only if $\hat{u}$ is a successor node of $\hat{w}$.
\end{enumerate}

For a node $\hat{u}$ of $T^c_G$ let $$V(\hat{u}) = \{ \vertex{\hat{v}} \,|\, \hat{v} \text{ is a leaf of }T^c_G(\hat{u})\}.$$ A vertex $v$ of $G$ which is not in $V(\hat{u})$ is either adjacent to all vertices of $V(\hat{u})$ or to none of them. This depends on whether the first common predecessor of $\hat{u}$ and $\node{v}$ in $T$ is a join node or a union node, respectively.

The following algorithm computes a minimum vertex set that is a vertex cover for $\SR{G}$ which contains all vertices that are maximally distant from a vertex $w$ in $G$. We assume that $\text{vc}(\hat{u})$ is already computed for each node $\hat{u}$ of $T^c_{\SR{G}}$ by Algorithm\,1.

\begin{algorithm}[ht]
\SetKwProg{Fn}{Algorithm\,2\,}{$(T^c_G,T^c_{\SR{G}},w)$}{}
\Fn{}{
	$\hat{w} \leftarrow \node{w}$; \\
	\If{$(\hat{w} = \hat{r})$}{
		\Return 0;
	}
	{\rm\bf Let} $\hat{v}$ be the predecessor node of $\hat{w}$ in $T^c_G$; \\
	$\text{h}(\hat{v}) \leftarrow \text{n}({\hat{v})} - 1$; \\
	\While{$(\hat{v} \not= \hat{r})$}{
		{\rm\bf Let} $\hat{u} \leftarrow \hat{v}$; \\
		{\rm\bf Let} $\hat{v}$ be the predecessor node of $\hat{u}$ in $T^c_{\SR{G}}$; \\
		{\rm\bf Let} $\hat{u}_1,\ldots,\hat{u}_k$ be the successor nodes of $\hat{v}$ in $T^c_{\SR{G}}$ without node $\hat{u}$; \\
		\If{$(\hat{v}$ is a join node in $T^c_{\SR{G}})$}{
			\tcp{\it $\hat{v}$ is a union node in $T^c_G$}
			$\text{h}(\hat{v}) \leftarrow \text{h}(\hat{u}) + \text{n}({\hat{u}_1}) + \cdots + \text{n}({\hat{u}_k})$;
		}
		\Else{
			\tcp{\it $\hat{v}$ is a union node in $T^c_{\SR{G}}$}
			\tcp{\it $\hat{v}$ is a join node in $T^c_G$}
			$\text{h}(\hat{v}) \leftarrow \text{h}(\hat{u}) + \text{vc}({\hat{u}_1}) + \cdots + \text{vc}({\hat{u}_k})$;
		}
	}
	\Return $\text{h}(\hat{r})$;
}
\end{algorithm}

\begin{figure}[ht]
\center
\includegraphics[width=360pt]{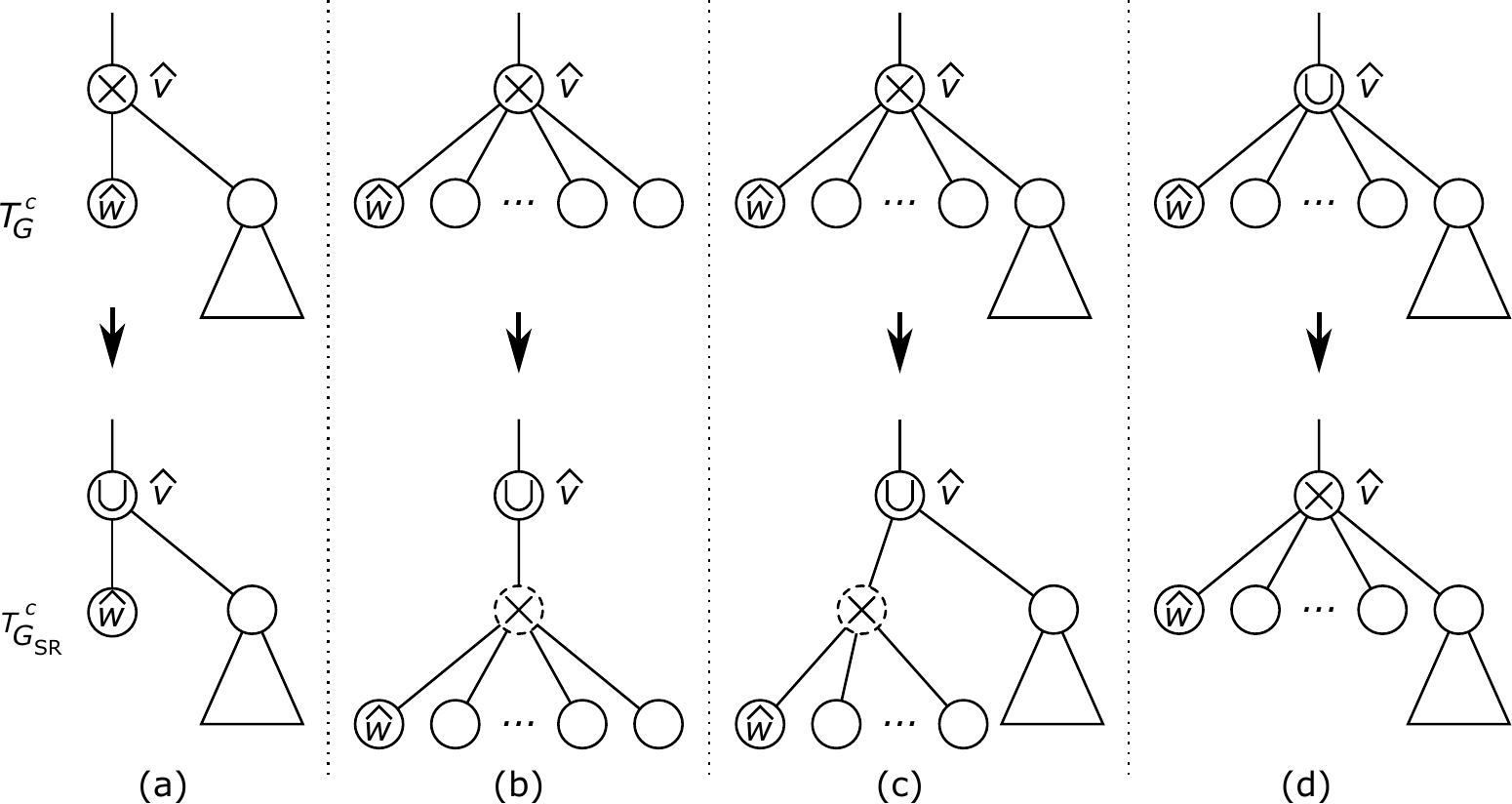}
\caption{The 4 cases when node $\hat{v}$ is the predecessor node of $\hat{w}$ in $T^c_G$.}
\label{figure11}
\end{figure}

Algorithm\,2 initially sets variable $\text{h}(\hat{v})$ to $\text{n}({\hat{v})} - 1$ for the predecessor node $\hat{v}$ of $\hat{w}=\node{w}$ in $T^c_G$. To explain the correctness of this instruction, we distinguish between the 4 cases shown in \Cref{figure11}. In the cases (a), (b), and (c) node $\hat{v}$ is the predecessor node of $\hat{w}$ in $T^c_G$. In these cases, all vertices of $V(\hat{v}) \setminus w$ are maximally distant from $w$ in $G$, based on the second consideration above. One vertex is subtracted here, since vertex $w$ is not maximally distant from $w$ itself if $G$ is connected and has at least two vertices. In case (d), node $\hat{v}$ is a union node in $T^c_G$ and thus all vertices of $V(\hat{v}) \setminus w$ are maximal distant from $w$ based on the first consideration above. If $\hat{v}$ is a node further up on the path to the root $\hat{r}$, we only have to distinguish whether $\hat{v}$ is a join or a union node in $T^c_{\SR{G}}$. If $\hat{v}$ is a join node in $T^c_{\SR{G}}$, then $\hat{v}$ is a union node in $T^c_G$ and all vertices of $V(\hat{v}) \setminus V(\hat{u})$ are maximally distant from $w$. If $\hat{v}$ is a union node in $T^c_{\SR{G}}$, then the minimum vertex covers of the subgraphs induced by the vertex sets $V(\hat{u_i})$ of the successor nodes $\hat{u_i}$ of $\hat{v}$ without node $\hat{u}$ have to be merged. In this case, node $\hat{v}$ is a join node in $T^c_{G}$ and the vertices of $V(\hat{v}) \setminus V(\hat{u})$ are not maximal distant from $w$.

%%%%%%%%%%%%%%%%%%%%%%%%%%%%%%%%%%%%%%%%%%%%%%%%%%%%%%%%%%%%
% Figure 11
%%%%%%%%%%%%%%%%%%%%%%%%%%%%%%%%%%%%%%%%%%%%%%%%%%%%%%%%%%%%

\begin{figure}[ht]
\center
\includegraphics[width=240pt]{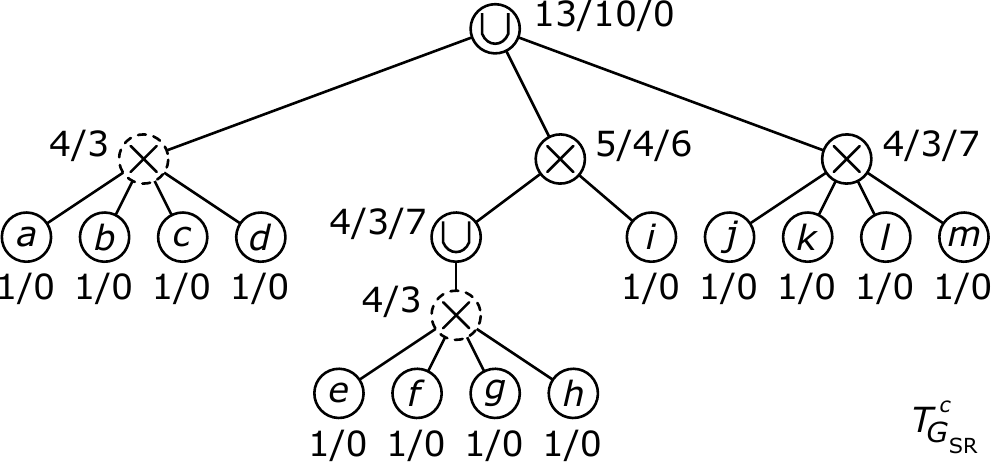}
\caption{A canonical co-tree $T^c_{\SR{G}}$ for a co-graph $\SR{G}$. The inner nodes $\hat{u}$ which also exist in $T^c_G$ are labelled $\text{n}(\hat{u})/\text{vc}(\hat{u})/\text{m}(\hat{u})$ by Algorithm\,3. The leafs and the twin-join nodes are labelled only by $\text{n}(\hat{u})/\text{vc}(\hat{u})$. For example, \\[12pt]
\centerline{$\begin{array}{lllll}
\XVC{\SR{G}}{\MD{G}{\capfont{a}}} & = & 13 -1 + 0 & = & 12,\\
\XVC{\SR{G}}{\MD{G}{\capfont{e}}} & = & 4 -1 + 7 & = & 10,\text{ and}\\
\XVC{\SR{G}}{\MD{G}{\capfont{i}}} & = & 5 -1 + 6 & = & 10.\\
\end{array}$}
}
\label{figure10}
\end{figure}

As in Section \Cref{section_cycles}, let
$$h_i = \vert\XVC{\SR{G(\hat{u})} \setminus \{\sv{\hat{v}_1},\ldots,\sv{\hat{v}_k},\sv{\hat{w}}\}}{\MD{G(\hat{u})}{\sv{\hat{v}_i}}\setminus \{\sv{\hat{v}_1},\ldots,\sv{\hat{v}_k},\sv{\hat{w}}\}}\vert$$
and $i_{min}$ be the index $i$ from $\{1,\ldots,k\}$ such that
$$h_i \, - \, \vert\XVC{\SR{\widetilde{G}(\hat{u_i})} \setminus\{\sv{\hat{v}_i}\}}{\MD{\widetilde{G}(\hat{u_i})}{\sv{\hat{v}_i}}}\vert$$
is minimal. To determine $i_{\min}$, we have to compute $h_i$ for all $i$ in total linear time. To achieve this, we calculate the increments of the values $\text{h}(\hat{u})$ at the inner nodes $\hat{u}$ of $T^c_G$ along the path to the root $\hat{r}$ top-down in a preprocessing phase. Algorithm\,3 computes all these increments, denoted by $\text{m}(\hat{u})$, in total linear time, see also \Cref{figure10}.

\begin{algorithm}[ht]
\SetKwProg{Fn}{Algorithm\,3\,}{$(T^c_G,T^c_{\SR{G}})$}{}
\Fn{}{
	\For{$($each inner node $\hat{u}$ of $T^c_G)$}{
		\tcp{\it $-1$ means undefined}
		$\text{m}(\hat{u}) \leftarrow -1$;
	}
	$\text{m}(\hat{r}) \leftarrow 0$; \\
	\While{
	$\left(
	\begin{array}{l}
	\text{there is an inner node } \hat{u} \text{ in } T^c_G \text{ with predecessor node } \hat{v} \\
	\text{such that } \text{\rm vc}({\hat{u}}) = -1 \text{ and } \text{\rm vc}({\hat{v}}) \geq 0
	\end{array}
	\right)$}{
		\ \\
		{\rm\bf Let} $\hat{u}_1,\ldots,\hat{u}_k$ be the successor nodes of $\hat{v}$ in $T^c_{\SR{G}}$ without node $\hat{u}$; \\
		\If{$(\hat{v}$ is a join node in $T^c_{\SR{G}})$}{
			\tcp{\it $\hat{v}$ is a union node in $T^c_G$}
			$\text{m}(\hat{u}) \leftarrow \text{m}(\hat{v}) + \text{n}({\hat{u}_1}) + \cdots + \text{n}({\hat{u}_k})$;
		}
		\Else{
			\tcp{\it $\hat{v}$ is a union node in $T^c_{\SR{G}}$}
			\tcp{\it $\hat{v}$ is a join node in $T^c_G$}
			$\text{m}(\hat{u}) \leftarrow \text{m}(\hat{v}) + \text{vc}({\hat{u}_1}) + \cdots + \text{vc}({\hat{u}_k})$;
		}
	}
}
\end{algorithm}

After the pre-processing by Algorithm\,3, the size $h_i$ of $$\XVC {\SR{G}(\hat{u}))} {\MD{G(\hat{u_i})}{\sv{\hat{v}_i}}}, \, 1\leq i \leq k,$$ is computable in time $\mathcal{O}(1)$ as follows. If $\hat{u}$ is the predecessor node of $\hat{u}_i$ in $T^c_{\widetilde{G}}$, then $h_i = \text{n}({\hat{u})} - 1 + \text{m}(\hat{u})$, see \Cref{figure10}.

%%%%%%%%%%%%%%%%%%%%%%%%%%%%%%%%%%%%%%%%%%%%%%%%%%%%%%%%%%%%
% Figure 10
%%%%%%%%%%%%%%%%%%%%%%%%%%%%%%%%%%%%%%%%%%%%%%%%%%%%%%%%%%%%

\begin{figure}[ht]
\center
\includegraphics[width=240pt]{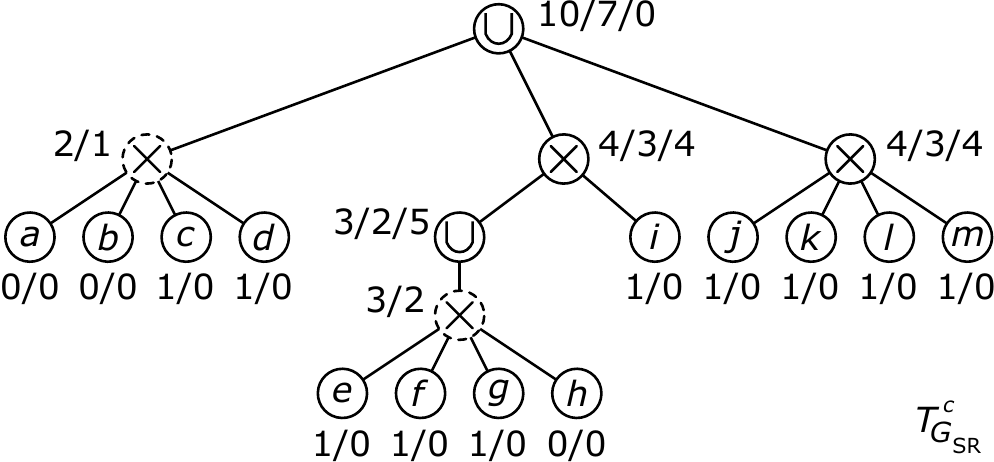}
\caption{A canonical co-tree $T^c_{\SR{G}}$ for a co-graph $\SR{G}$. The inner nodes $\hat{u}$ which also exist in $T^c_G$ are labelled $\text{n}(\hat{u})/\text{vc}(\hat{u})/\text{m}(\hat{u})$ by Algorithm\,4. The leafs and the twin-join nodes are labelled only by $\text{n}(\hat{u})/\text{vc}(\hat{u})$. Here, the vertices \capfont{a}, \capfont{b}, and \capfont{h} are left out by setting $\text{n}(\capfont{a})$, $\text{n}(\capfont{b})$, and $\text{n}(\capfont{h})$ to zero. For example, \\[12pt]
\centerline{$\begin{array}{lllll}
\XVC{\SR{G} \setminus \{\capfont{a},\capfont{b},\capfont{h}\}}{\MD{G}{\capfont{a}} \setminus \{\capfont{a},\capfont{b},\capfont{h}\}} & = & 10 -0 + 0 & = & 10,\\
\XVC{\SR{G} \setminus \{\capfont{a},\capfont{b},\capfont{h}\}}{\MD{G}{\capfont{e}} \setminus \{\capfont{a},\capfont{b},\capfont{h}\}} & = & 3 -1 + 5 & = & 7,\text{ and}\\
\XVC{\SR{G} \setminus \{\capfont{a},\capfont{b},\capfont{h}\}}{\MD{G}{\capfont{h}} \setminus \{\capfont{a},\capfont{b},\capfont{h}\}} & = & 3 -0 + 5 & = & 8.\\
\end{array}$}
}
\label{figure9}
\end{figure}

Finally, we need the size of the set $$h_i = \vert\XVC{\SR{G(\hat{u})} \setminus \{\sv{\hat{v}_1},\ldots,\sv{\hat{v}_k},\sv{\hat{w}}\}}{\MD{G(\hat{u})}{x_i}\setminus \{\sv{\hat{v}_1},\ldots,\sv{\hat{v}_k},\sv{\hat{w}}\}}\vert$$ where the nodes $\sv{\hat{v}_1},\ldots,\sv{\hat{v}_k},\sv{\hat{w}}$ are left out from $\SR{G(\hat{u})}$. This can be achieved by initially setting the $\text{n}$-values of these nodes to zero and not to 1, see \Cref{figure9}. Here, two cases must be distinguished for the calculation of $h_i$. If vertex $u$ does not belong to the set of vertices that are excluded, then the calculation of $h_i$ is as before $h_i = \text{n}(\hat{u})-1+\text{m}(\hat{u})$. However, if node $u$ has been excluded, the computation of $h_i$ is $h_i = \text{n}(\hat{u})+\text{m}(\hat{u})$. Both cases can be covered by $$h_i = \text{n}(\hat{u})-\text{n}(\hat{w})+\text{m}(\hat{u}).$$

Since index $i_{\min}$ is computable in linear time, the sizes of
\begin{itemize}
\item
$\MVC{\SR{\widetilde{G}(\hat{u})}}$,
\item
$\MVC{\SR{\widetilde{G}(\hat{u})}\setminus\{\sv{\hat{w}}\}}$, and
\item
$\XVC{\SR{\widetilde{G}(\hat{u})}\setminus\{\sv{\hat{w}}\}}{\MD{\widetilde{G}(\hat{u})}{\sv{\hat{w}}}}$
\end{itemize}
are computable in linear time by \Cref{lemma7,lemma8,lemma9}.

%%%%%%%%%%%%%%%%%%%%%%%%%%%%%%%%%%%%%%%%%%%%%%%%%%%%%%%%%%%%
% Figure 9
%%%%%%%%%%%%%%%%%%%%%%%%%%%%%%%%%%%%%%%%%%%%%%%%%%%%%%%%%%%%

%%%%%%%%%%%%%%%%%%%%%%%%%%%%%%%%%%%%%%%%%%%%%%%%%%%%%%%%%%%%
% Theorem 4
%%%%%%%%%%%%%%%%%%%%%%%%%%%%%%%%%%%%%%%%%%%%%%%%%%%%%%%%%%%%

\begin{theorem}
\label{theorem4}
A minimum strong resolving set for a graph $G=(V,E)$, in that each biconnected component is a co-graph, can be computed in time ${\mathcal O}(|V| + |E|)$.
\end{theorem}

Ollermann and Peters-Fransen showed in \cite{OP07} that the size of a strong resolving set can be computed in polynomial time for distance hereditary graphs. Graphs in which the biconnected components are co-graphs are distance hereditary, but our solution presented here runs in linear time.

%%%%%%%%%%%%%%%%%%%%%%%%%%%%%%%%%%%%%%%%%%%%%%%%%%%%%%%%%%%%
% Section 8
%%%%%%%%%%%%%%%%%%%%%%%%%%%%%%%%%%%%%%%%%%%%%%%%%%%%%%%%%%%%

\section{Conclusion}
\label{section8}

In this paper we have shown that the efficient computation of a strong resolving set for a graph $G$ essentially depends on the efficient computation of strong resolving sets for its biconnected components. If a minimum strong resolving set can be computed for a biconnected graph in polynomial time, it is generally also possible to compute minimum strong resolving sets in polynomial time, which additionally contain the vertices that are maximally distant from other vertices.

We have given three examples for which it is possible to compute the required assumptions in linear time. From this it could be concluded that the computation of minimum strong resolving sets for graphs is possible in linear time if the biconnected components are circles or co-graphs. It would be interesting to know for which other more complex graph classes this concept is applicable.

A generalization of the procedure for directed graphs and directed strong resolving sets, see for example \cite{SW21}, as well as a generalization of the composition of two graphs over several vertices that are all connected to each other are also interesting challenges.

%%%%%%%%%%%%%%%%%%%%%%%%%%%%%%%%%%%%%%%%%%%%%%%%%%%%%%%%%%%%%%%%%%%%%%%%%%
\newcommand{\etalchar}[1]{$^{#1}$}

%%%%%%%%%%%%%%%%%%%%%%%%%%%%%%%%%%%%%%%%%%%%%%%%%%%%%%%%%%%%%%%%%%%%%%%%%%

%%%%%%%%%%%%%%%%%%%%%%%%%%%%%%%%%%%%%%%%%%%%%%%%%%%%%%%%%%%%%%%%%%%%%%%%%%
\end{document}